# Transport Properties of operational gas mixtures used at LHC

Y.Assran[*]

Suez Canal University, Suez, Egypt

and

A.Sharma

CERN CH1211 Geneve, Switzerland

**Abstract**

This report summarizes some useful data on the transport characteristics of gas mixtures which are required for detection of charged particles in gas detectors. We try to replace Freon used for RPC detector in the CMS experiment with another gas while maintaining the good properties of the Freon gas mixture unchanged. We try to switch to freonless gas mixture because Freon is not a green gas, it is very expensive and its availability is decreasing. Noble gases like Ar, He, Ne and Xe (with some quenchers like carbon dioxide, methane, ethane and isobutene) are investigated. Transport parameters like drift velocity, diffusion, Townsend coefficient, attachment coefficient and Lorentz angle are computed using Garfield software for different gas mixtures and compared with experimental data.

## 1    Introduction

Four main experiments have been designed and constructed for the Large Hadron Collider (LHC) machine: ATLAS, CMS, LHCb and ALICE. ATLAS and CMS are large general-purpose experiments, LHCb will study *b*-quark systems, produced predominantly in the forward direction, and ALICE is designed specifically for studies of heavy-ion collisions. Figures 1-4 show the schemes of the four detectors. The muon systems in all these experiments are large-area gas-based detectors (several thousand $m^2$ of multilayer chambers each in ATLAS and CMS). The chambers are divided into two sets, one intended for precise measurements of muon tracks and the other dedicated to triggering on muons. In this paper we will briefly summarize the main characterization of these detectors and provide a summary of the transport properties of the operation gas mixtures.

## 2    Gas detectors at LHC
## 2-1  Gas detectors at CMS

The CMS muon detector uses three types of gaseous particle detectors for muon identification [1]: Drift Tubes (DT) in the central barrel region, Cathode Strip Chambers (CSC) in the endcap regions and Resistive Parallel Plate Chambers (RPC) in both the barrel and endcaps. The DT and CSC detectors are used to obtain a precise measurement of the position and thus the momentum of the muons, whereas the RPC chambers are dedicated to providing fast information for the Level-1 trigger. In the following different types of detector and their use in the CMS Muon Detector will be defined in some detail.

---

[*] Corresponding author, Presently at CERN, Geneve, Switzerland,

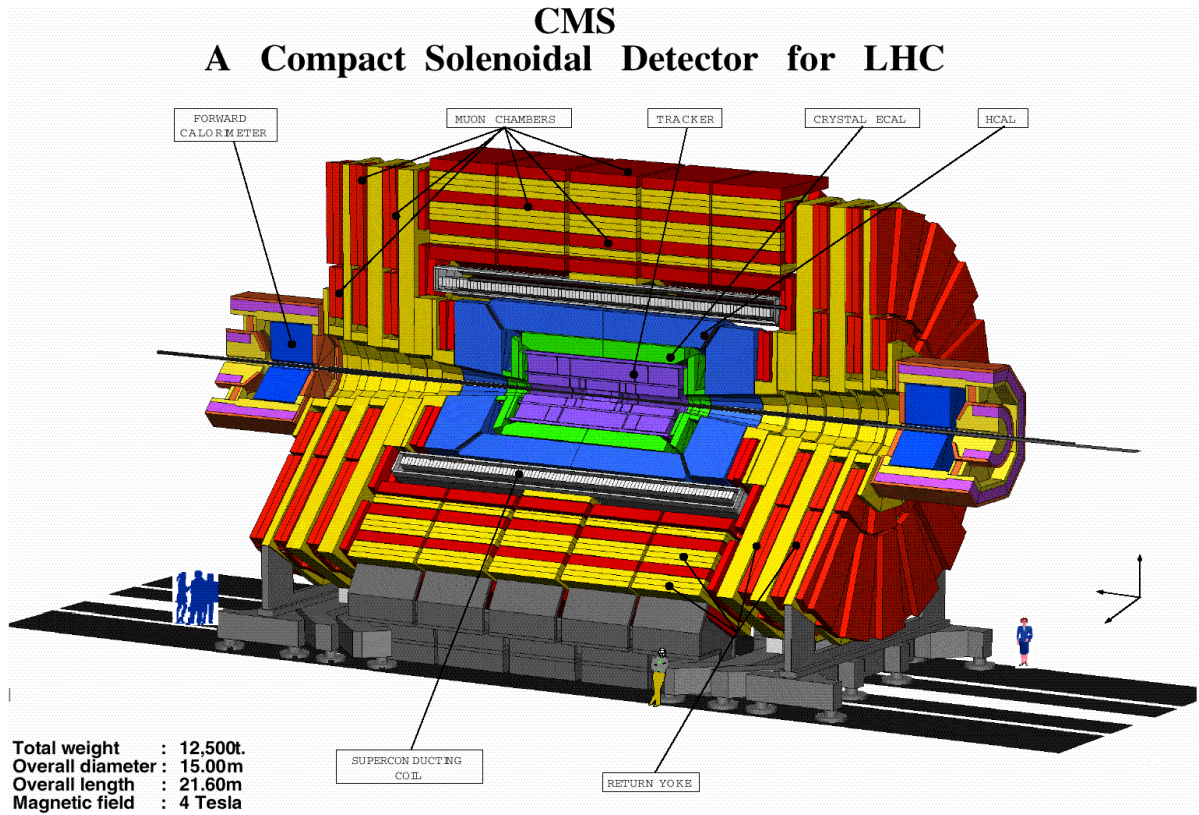

Figure 1. The CMS detector.

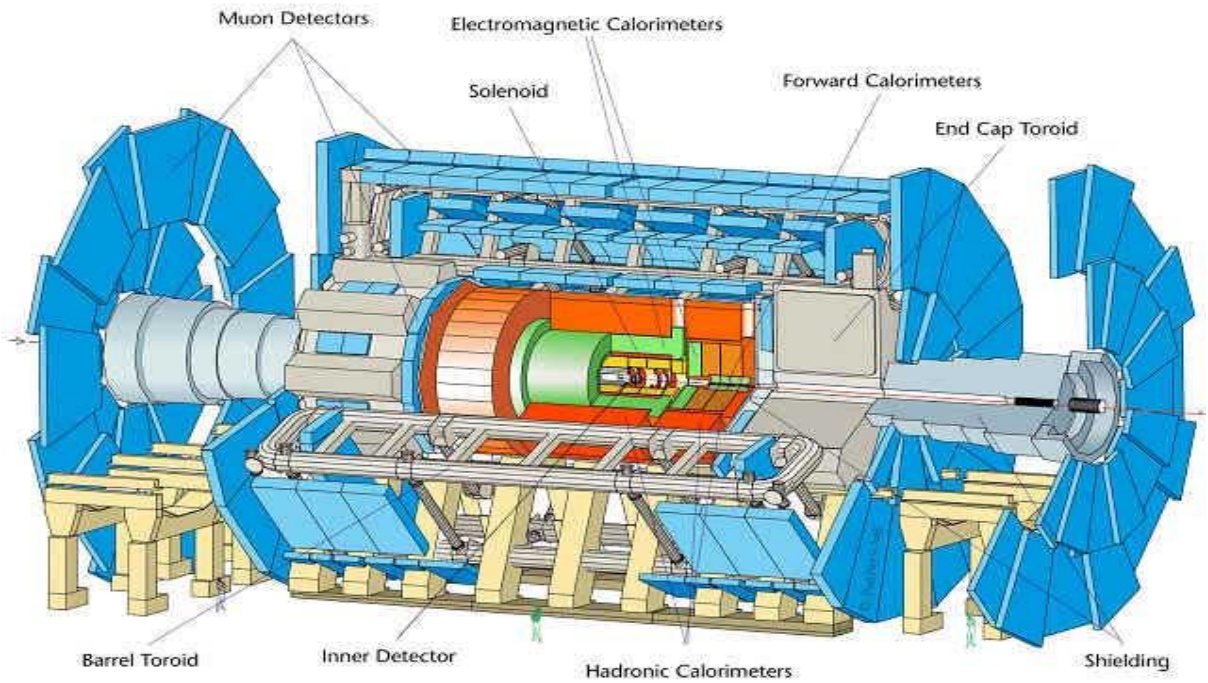

Figure 2. The ATLAS detector

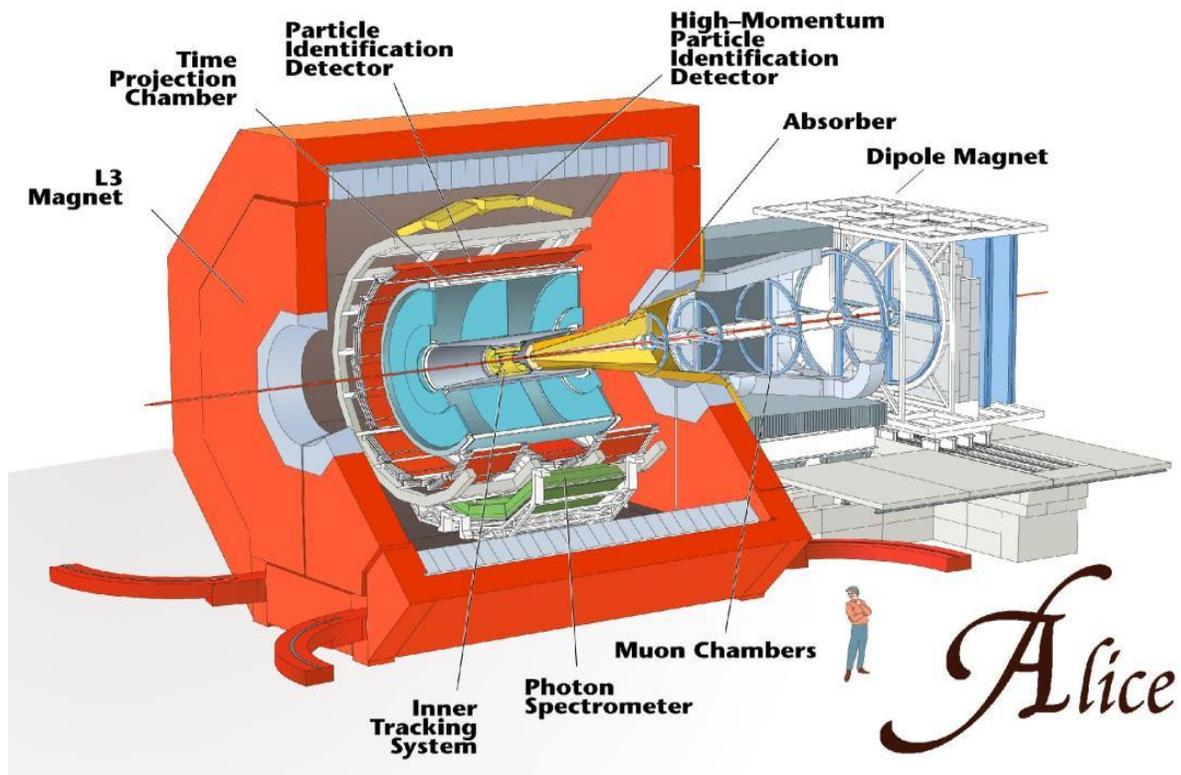

Figure 3. the ALICE detector.

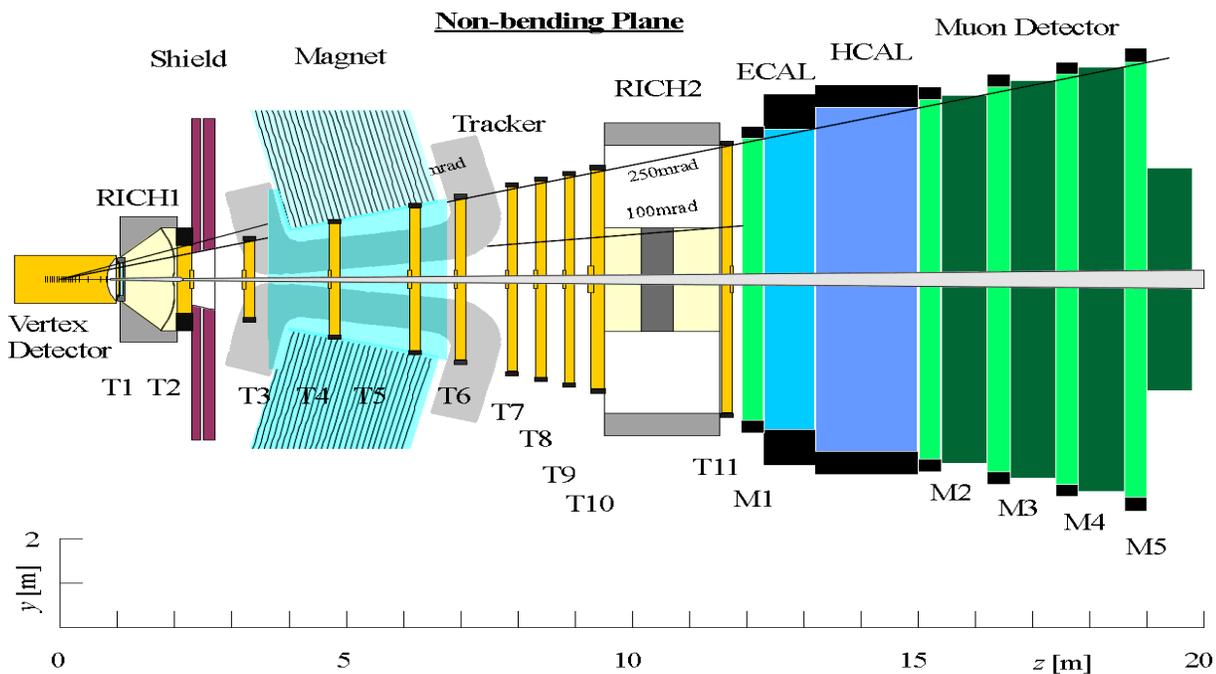

Figure 4. The LHCb detector

Drift Tubes are used in the Barrel where the magnetic field is guided and almost fully trapped by the iron plates of the Magnet Yoke. Each tube contains a wire with large pitch (4 cm), and the tubes are arranged in layers. Only the signals from the wires are recorded - resulting in a moderate number of electronic channels needed to read out the detectors. When a charged particle passes through the DT, it forms electron-ion pairs inside the gas mixture (Ar 85% - $CO_2$ 15%). The electrons follow the electric field and end up at the positively charged wire.

A DT layer is put together by gluing an aluminium plate to a set of parallel aluminium-I beams. The wires are stretched, held by appropriate end plugs, and the layer is closed by another aluminium plate. Each DT consists of twelve aluminum layers arranged in three groups of four. The middle group measure the coordinate along the beam direction and the two outside groups measure the perpendicular coordinate. Copper strips are previously glued to the Al plates in front of the wire to better shape the electrostatic field. A full-size final prototype of a DT chamber is shown below. The chamber is 2m x 2.5m in size. The largest DT chambers to be used in CMS will have dimensions of 4m x 2.5m in size.

Cathode strip chambers are used in the end cap regions where the magnetic field is very intense (up to several Tesla) and very inhomogeneous to provide precise space and time information. It consist of positively charged wires (anode) crossed with negatively charged copper strips (cathode) within a gas volume (Ar 40% - $CO_2$ 50% - $CF_4$ 10%)**.** When a muon passes the chamber it ionizes the gas and forms electron-ion pairs inside it. The electrons go to the anode wires creating an avalanche, the positive ions move toward the copper cathode strips, inducing a charge pulse in the strip. Two position coordinates for each particle are recorded. The wires give the radial coordinate whereas the strips measure Φ.

In addition to providing precise space and time information, the closely spaced wires (2mm) make the CSC a fast detector suitable for triggering. CSC modules containing six layers provide both robust pattern recognition for rejection of non-muon backgrounds and also efficient matching of external muon tracks to internal track segments.

Resistive plate chambers are gaseous detectors which aim to provide a muon trigger in both the barrel and end cap regions. It has been proposed as a suitable solution to build first level muon trigger because of its fast response [2], and good time resolution, flexibility in segmentation, robustness and the relatively low cost of production.

The RPC is a detector utilizing a constant and uniform electric field produced by two resistive parallel electrode plates. The 2mm thick gap between these two plates was filled with a gas (Freon 96.2% + isobutane 3.5% + $SF_6$ 0.3%) of high absorption coefficient to ultraviolet light. When charged particles pass through the gas, the gas atoms will ionize and form electron-ion pairs. The electron goes to the positive electrode and discharge is originated by the electric field. The discharge is prevented from propagation through the whole gas by the high resistivity of the electrodes.

The resistive plate chambers used in the CMS experiment consist of two layers of gas gaps with a sheet of copper readout strips sandwiched between them. Gas gap is made of two sheets of high resistive Bakelite which acts as the electrode. Within the resistive plate chamber the electric field is uniform. The sheet of the readout strips is put in the chamber, centered on the bottom gap. There are 32 strips running the length of the chamber but they are broken up lengthwise into three sections to form appropriate trigger towers.

## 2-2 Gas detectors in ATLAS

The ATLAS experiment uses gas detectors in both the inner detector and muon spectrometer [3]. In inner detector of ATLAS uses Transition radiation Detectors (TRT) to resolve the inevitable ambiguities caused by overlapping tracks, secondary interactions, detector inefficiencies and alignment errors. In muon the spectrometer, ATLAS uses Monitored Drift Tube (MDT) and Cathode Strip Chamber (CSC) for precision measurements; Resistive Plate Chamber (RPC) and Thin Gap Chamber (TGC) for triggering.

TRT provides tracking and contributes to the electron identification over the whole inner detector rapidity range. Its pattern recognition capability is strong due to the large number of measurement points (>36 per track), which will be combined to perform momentum measurement together with the semi-conducting tracking (SCT) detectors. TRT can also provide a stand-alone momentum measurement, but with a lower precision than the whole inner detector. The detector will be built in three deferent blocks – two end-cap TRTs with radial straws and one barrel TRT with axially-oriented straws. So, the barrel TRT measures R $\phi$ while the end-cap TRT measures $\phi$ and z directly and R indirectly through the particle entrance and exit positions in the detector.

The active element of TRT is the Straw which is made of two polyimide films, with 2000A aluminum covered with 4µm carbon, is wound to form the 4mm diameter straws with a wall thickness of 60 µm. these straws are then reinforced by gluing four carbon fibers along the straw wall, giving stable mechanical properties. A 50 µm gold-plated copper/beryllium wire is situated along the straw axis. The chamber gas is chosen to be $70\%Xe + 20\%CF_4 + 10\% CO_2$. The straw operates with a high voltage 1780 V giving a gas gain of $2.5*10^4$. The maximum drift time is 38 ns for this gas mixture in a 2T magnetic field.

The MDTs are employed in both barrel and end cap regions. In the endcap regions it is used up to rapidities close to $\eta = 2$, where the rate approaches 200 Hz/cm$^2$ and occupancies are at the 7-8% level. An MDT is built from two multilayers of high-pressure drift tubes mounted to either side of a support structure. Each multilayer consists of three or four layers of tubes. The support structure uses a minimum of material and provides precise positioning of the two multilayers and alignment components of the system. The possible distortion of the chambers is monitored with four optical straightness monitors. A relatively fast drift gas with maximum drift time <600 ns is needed to limit the occupancy but conflict the requirement of small Lorentz angle. The tubes must be operates at a low gas gain of $< 5.10^4$ to limit the level of streamers to less than 1%, and to minimize chamber ageing. Nonflammability of the mixtures is also required. Several gas mixtures have been studied, such as Ar-$C_2H_6$-$CO_2$-$N_2$ (86-5-4-5) or Ar-$C_2H_6$-$N_2$ (90:5:5).

The second precision chamber is the CSC which is used at higher rapidity because of the need to increase the granularity, also for considerations on chamber ageing. CSC is a multiwire proportional chamber with a symmetric cell in which the anode-cathode distance equal the anode wire spacing which is 2.5mm, the cathode is segmented in strips with a readout pitch of 5mm. the transverse coordinate can be determined by segmented the second cathode in strips parallel to the anode wires or, by reading the anode wires. The operating gas is a mixture of $Ar$-$CO_2$-$CF_4$ (30:50:20). This gas combines a high drift velocity v>60 μm/ns with a small Lorentz angle. The high drift velocity and the narrow anode wire spacing result in a short total drift time (<25 ns).

The trigger system employs two types of detectors, (RPC) and (TGC). The combination optimizes the cost of the total detector, matching the rate capability and the spatial resolution to the requirement. The RPC is a gaseous detector which was already described in the previous section but here it uses two sets of orthogonal strips for two-dimensional readout of the particle position. The conventional operating mode of the RPC uses a gas mixture of argon-butane with a few per cent of Freon. Using this mode, large gas amplification limits the rate capability of the detector to about 100 $Hz/cm^2$. A higher rate capability can be obtained with 100 times smaller gas gain, using a gas mixture of $Ar$-butane-$SF_6$ (74:8:18). The rate capability of RPC operating with low gas amplification was found to be one order of magnitude higher than the conventional mode.

The second trigger chamber is the TGC which uses 50μm wires with 2mm pitch, sandwiched between two graphite cathodes at a distance of 1.6 mm from the anode plane. Behind the graphite cathode planes are strips to perform a capacitive readout. The most important characteristics of the TGC are : fast signal with r.m.s time resolutions < 5ns, rate capability up to 150kHz/$cm^2$, saturated mode operation, small sensitivity to magnetic field ,and strips and wire readout in almost any geometrical form. The time resolutions were obtained with a mixture of $CO_2$ − n-pentane (55:45). Other mixtures have been tried such as $CF_4$ − isobutene (80:20), which gives a timing resolution better than 4 ns for a single layer. However, the operating conditions for this type of gas need further investigation.

## 2-3 Gas detectors at LHCb

The LHCb muon system consists of five stations. Multiwire proportional chambers (with a mixture of Ar 40% + $CO_2$ 55% + $CF_4$ 5%) are used throughout, except for the innermost region closest to the beamline of the first station. This station is placed in front of the calorimeters and represents a significant challenge in terms of material budget, space constraints, rate capability and radiation tolerance. The innermost region of this station, where the particle rates are highest, is equipped with triple-GEM (gas electron multiplier) detectors [4]with pad readout that are particularly suited for tracking in a high particle rate environment. Prototype tests have shown that the fast $Ar/CO_2/CF_4$ (45:15:40) gas mixture allowed to achieve a time resolution better than 3 ns, to be compared with the time resolution of ~10 ns obtained with the standard $Ar/CO_2$ (70:30) mixture [5] . For outer tracker, LHCB use straw tubes with gas mixture of $Ar/CO_2/CF_4$ (70:10:15). Like CMS and ATLAS, LHCB use RPC chamber for triggering purposes with gas mixture of $C_2H_2F_4/iC_4H_{10}/SF_6$ (95:4:1).

## 2-4 Gas detectors at ALICE

The ALICE experiment has a central barrel, housed in the L3 magnet, covering in pseudorapidity the range $-0.9 \leq \eta \leq 0.9$ with complete azimuthal coverage [6]. This central barrel comprises an inner tracking system of Silicon detectors (ITS), a large time projection chamber (TPC), a transition radiation detector (TRD) and a time-of-flight array (TOF). The main goal of the TRD is to provide electron identification in the central barrel at momenta in excess of 1 GeV/$c$ where the pion rejection via energy loss measurement in the TPC is no longer sufficient. The gas mixture used in the TRD chambers is $Xe/CO_2$ in a ratio of 85/15 while that used in TPC chamber is $Ne/CO_2$ with the ratio of (90:10) and that used for TOF chambers is $C_2H_2F_4/iC_4H_{10}/SF_6$ with the ratio of (90:5:5). ALICE uses also RPC chambers for muon spectrometer, the gas mixture used in those chambers is $C_2H_2F_4/iC_4H_{10}/SF_6$ with the ratio of (89.7:10:0.3).

## 3 Primary and total ionization

When a charged particle passing through a gaseous medium, it ionizes atoms and molecules of the gas and free electrons and ions are produced in amounts depending on the atomic number, density and ionization potential of the gas, and on the charge and energy of the incident particle. The created electrons can have sufficient energy to make further ionization and create secondary electron-ion pairs. The former process is called primary ionization and the number of primary electron-ion pairs per cm is called $N_p$. the sum of the two processes is total ionization and the total number of electron-ion pairs per cm is called $N_t$. $N_p$ and $N_t$ are characteristic of a given gas or gas mixture. The spatial resolution of a chamber depends on various factors, $N_p$ and $N_t$ are the most important ones. These numbers have been measured and computed for a variety of gases [7-14]. Table 1 lists some values for $N_p$ and $N_t$ along with other general properties for some gases and gas mixtures. All numbers are for normal temperature and pressure (NTP). For gas mixtures a weighted average for $N_p$ and $N_t$ may be computed. For equal gas density, the number of primaries is larger for gases whose molecules comprise light elements. The numbers given are for minimum ionizing particles; for lower energy particles the larger ionization density helps improving accuracy.

| Gas | Ratio | Density$\times 10^{-3}$ (g/cm$^3$) | Radiation Length (m) | $N_p$ (cm$^{-1}$) | $N_t$ (cm$^{-1}$) |
|---|---|---|---|---|---|
| He | | o.178 | 5299 | 4.8 | 8 |
| Ar | | 1.782 | 110 | 24.3 | 94 |
| Ne | | 0.9 | 345 | 12 | 43 |
| Xe | | 5.86 | 15 | 44 | 307 |
| CF$_4$ | | 3.93 | 92.4 | 51 | 100 |
| DME | | 2.2 | 222 | 60 | 160 |
| CO$_2$ | | 1.98 | 183 | 35.5 | 91 |
| CH$_4$ | | 0.71 | 646 | 26.5 | 53 |
| C$_2$H$_6$ | | 1.34 | 340 | 41 | 111 |
| i-C$_4$H$_{10}$ | | 2.59 | 169 | 84 | 195 |
| Ar-CH$_4$ | 90-10 | 1.67 | 120 | 24.5 | 90 |
| | 80-20 | 1.57 | 132 | 24.7 | 85.8 |
| | 70-30 | 1.46 | 147 | 25 | 82 |
| Ar-C$_2$H$_6$ | 90-10 | 1.74 | 118 | 26 | 95.7 |
| | 80-20 | 1.69 | 127 | 27.6 | 97.4 |
| | 70-30 | 1.65 | 138 | 29 | 99.1 |
| Ar-iC$_4$H$_{10}$ | 90-10 | 1.86 | 114 | 30.27 | 104.1 |
| | 80-20 | 1.94 | 118 | 36.24 | 114.2 |
| | 70-30 | 2 | 122.8 | 42.21 | 124.3 |
| Ar-CO$_2$ | 90-10 | 1.8 | 114.5 | 25.42 | 93.7 |
| | 80-20 | 1.82 | 119.5 | 26.54 | 93.4 |
| | 70-30 | 1.84 | 124.9 | 27.66 | 93.1 |
| He-CH$_4$ | 90-10 | 0.237 | 3087 | 7 | 12.5 |
| | 80-20 | 0.285 | 2178 | 9.1 | 17 |
| | 70-30 | 0.355 | 1683 | 11.3 | 21.5 |
| He-C$_2$H$_6$ | 90-10 | 0.29 | 2155 | 8.42 | 18.3 |
| | 80-20 | 0.41 | 1353 | 12 | 28.6 |
| | 70-30 | 0.53 | 986 | 15.6 | 38.9 |
| He-iC$_4$H$_{10}$ | 90-10 | 0.42 | 1313 | 12.7 | 26.7 |
| | 80-20 | 0.66 | 749 | 20.6 | 45.4 |
| | 70-30 | 0.9 | 524 | 28.6 | 64.1 |
| He-CO$_2$ | 90-10 | 0.358 | 1396 | 8 | 16.3 |
| | 80-20 | 0.538 | 804 | 10.9 | 24.6 |
| | 70-30 | 0.719 | 565 | 14 | 32.9 |

Table 1: Parameters of some gas and gas mixtures.

| Gas | Ratio | Density*$10^{-3}$ (g/cm$^3$) | Radiation Length (m) | $N_p$ (cm$^{-1}$) | $N_t$ (cm$^{-1}$) |
|---|---|---|---|---|---|
| Ne-CH$_4$ | 90-10 | 0.881 | 361.8 | 13.45 | 44 |
| | 80-20 | 0.862 | 380.4 | 14.9 | 45 |
| | 70-30 | 0.843 | 401 | 16.35 | 46 |
| Ne-C$_2$H$_6$ | 90-10 | .0944 | 344 | 14.9 | 49.8 |
| | 80-20 | 0.988 | 343.9 | 17.8 | 56.6 |
| | 70-30 | 1.032 | 343.4 | 20.7 | 63.4 |
| Ne-iC$_4$H$_{10}$ | 90-10 | 1.06 | 312 | 19.2 | 58.2 |
| | 80-20 | 1.23 | 285 | 26.4 | 73.4 |
| | 70-30 | 1.4 | 262 | 33.6 | 88.6 |
| Ne-CO$_2$ | 90-10 | 1 | 317 | 14.35 | 47.8 |
| | 80-20 | 1.12 | 293 | 16.7 | 52.6 |
| | 70-30 | 1.22 | 272 | 19 | 57.4 |
| Xe-CH$_4$ | 90-10 | 5.34 | 16.6 | 42.25 | 281.6 |
| | 80-20 | 4.83 | 18.6 | 40.5 | 256.2 |
| | 70-30 | 4.31 | 21.2 | 38.75 | 230.8 |
| Xe-C$_2$H$_6$ | 90-10 | 5.4 | 16.6 | 43.7 | 287.4 |
| | 80-20 | 4.95 | 18.5 | 43.4 | 267.8 |
| | 70-30 | 4.5 | 21 | 43.1 | 248.2 |
| Xe-iC$_4$H$_{10}$ | 90-10 | 5.53 | 16.5 | 48 | 295.8 |
| | 80-20 | 5.2 | 18.3 | 52 | 284.6 |
| | 70-30 | 4.87 | 20.6 | 56 | 273.4 |
| Xe-CO$_2$ | 90-10 | 5.47 | 16.5 | 43.15 | 285.4 |
| | 80-20 | 5.1 | 18.4 | 42.3 | 263.8 |
| | 70-30 | 4.69 | 20.7 | 41.45 | 242.2 |

Table 1. (Continued)   Parameters of some gas and gas mixtures.

# 4 What should be the ideal gas mixture?

Particle physics experiments rely heavily on the detection of charged and neutral radiation with gaseous detectors. A suitable gas mixture enclosed within electrodes with an electric field between them permits the detection of charged particles. This mixture should fulfill a set of requirements depending upon the purpose it is used for. The mixture consists of primary gas in which ionization takes place. The most suitable gas for this are noble gases like helium,neon and argon, it should consist of another gas which is used as a quencher to prevent secondary effects like photon feedback and field emission in order to get a stable gas gain well separated from the noise of the electronics. Isobutane and methane can be used as possible quencher.

## 4-1 Tracking

For tracking at the high luminosity hadron collider like LHC, an operational gas mixture needs to have the following requisites: It has to be fast, so that an event can be unambiguously associated to its bunch crossing which leads to a compromise between having a high drift velocity and large primary ionization statistics. The drift velocity would ideally be saturated or have a small variation with modifications in electric and magnetic fields. The mixture needs to be well quenched. Fast ion mobility for quick clearance of positive ions to inhibit space charge effects which also helps in having small $E \otimes B$ effects.

Experiments at the upcoming high luminosity B-f and t-charm factories require a low mass gas mixture and a lot of work has been done in this direction. For low mass gas mixtures, used for detection of low momentum particles by minimizing multiple scattering, it is known that ethane, Isobutane and dimethyl ether (DME) are good quenchers with helium [15-20] , which help increasing also the number of primary and total ion pairs per cm for a given density.

Clearly financial constraints also need to be addressed in large gas systems, and a nonflammable, eco-friendly gas mixture is often a pre-requisite for safety.

## 4-2 Trigger

The most important characteristic of gas mixtures used for trigger purposes is that, it should be fast whether it has large primary ionization or not. As example CH4 can be used for this purpose because it is light and fast but we can't work at high voltage using pure CH4 so we use mixture of it and one of the noble gases like helium or argon.

# 5 Using Garfield for the calculation of transport parameters

Garfield [21] is a computer program for the detailed simulation of two- and three-dimensional drift chambers. It has interface to the Magboltz [22] program for the computation of electron transport properties in arbitrary gas mixtures. Garfield also has an interface with the Heed program to simulate ionization of gas molecules by particles traversing the chamber. Transport of particles, including diffusion, avalanches and current induction is treated in three dimensions irrespective of the technique used to compute the fields. In this work we use Garfield to calculate the transport parameters like drift velocity, longitudinal diffusion, transverse diffusion Townsend coefficient attachment coefficient and Lorentz angle, and then we compare our calculation with real data from different experiments.

**5-1 Drift Velocity.**

In the absence of electric field, electrons move randomly in all direction having an average thermal energy 3/2 KT. In presence of an electric field, the electrons start to drift along the field direction with mean drift velocity $v_d$ (the average distance covered by the drift electron per unit time). The drift velocity depends on the pressure, temperature and could be affected by the presence of pollutants like water or oxygen. Electronegative pollutant depletes the gas of electrons. Magboltz computes the drift velocity using Monte-Carlo techniques, tracing electrons at the microscopic level through numerous collisions with gas molecules.

**5-2 Diffusion**

Electrons and ions in a gas are subject only to an electric field and move on average along the electric field, but individual electrons deviate from the average due to scattering on the atoms of the gas. Scattering leads to variations in velocity, called longitudinal diffusion, and to lateral displacements, called transverse diffusion. The scattering process in each direction can to a good approximation be considered Gaussian on a microscopic scale. In cold gases like carbon-dioxide for example, the diffusion is small, while the drift velocity is low and unsaturated for values of electric fields common in gas detectors; this implies a non linear space time relation. Warm gases like argon for instance, have a higher diffusion; when mixed with polyatomic/organic gases having vibrational thresholds between 0.1 and 0.5 ev, diffusion is reduced in most cases, while the drift velocity is increased.

**5-3 Townsend coefficient**

The average distance an electron travels between ionizing collisions is called mean free path and its inverse is the number of ionizing collision per cm $\alpha$ (the first Townsend coefficient). This parameter determines the gas gain of the gas. If $n_o$ is the number of primary electron without amplification in uniform electric field, and n is the number of electrons after distance x under avalanche condition. So n is given by $n = n_o e^{\alpha x}$ and the gas gain G is given by $G = n_0/n = e^{\alpha x}$. The first Townsend Coefficient depends on the nature of the gas, the electric field and pressure.

**5-4 Lorentz angle**

Clearly due to the deflection effect exerted by a magnetic field perpendicular to the electric field and the motion of the electron, the electron moves in a helical trajectory resulting in a lowered drift velocity and transverse dispersion. Thus the arrival time of electrons in a proportional counter for example changes and the spread in the drift time increases. The angle which the drifting electron swarm makes with the electric field is defined as the Lorentz angle of the particular gas or gas mixture under consideration. This depends on both the electric field and the magnetic field. It is normally large at small electric fields but falls to smaller values for larger electric fields and is approximately linear with increasing magnetic field.

## 6 Simulation and experimental results

The various transport parameters have been computed and compared to the experimental results from literatures. Drift velocity of various gas mixtures is shown in figures (5-8), longitudinal and transverse diffusion are shown in figures (9-16), the Townsend and attachment coefficients are shown in figures (17-23), and Lorentz angle in figures (24-25). The mixtures considered for transport parameters are mainly Argon, Helium, Xenon and Neon since these are the mixtures typically used in the various experiments for tracking and triggering.

### 6-1 Drift velocity

Figure 5 shows the simulation results and experimental measurements for the Drift velocity of Argon based mixtures for different mixing ratios (start from 10% argon up to 90% with step of 10%); comparison with experimental results is made as far as possible. Most of the simulation results are corroborated by the experimental results to within 5-10%. For low electric field applications like the TPCs, Argon-Methane mixtures exhibit the highest drift velocity. For tracking and triggering applications which apply high electric field Argon-CO2 Mixtures are equally fast ~10cm/μsec.

Helium based mixtures which are shown in figures 6 exhibit relatively lower drift velocity as compared to argon based mixtures. Helium-CO2 mixtures are unsaturated while Helium-Isobutene mixtures are relatively saturated hence space-time correlations using Helium-Isobutene would be suitable.

In figure 7 are shown the drift velocity of Neon based mixtures. Neon mixtures are particularly useful for suppressing photon feedback and are used for example in the ALICE TPCs. Neon-CO2 mixtures are unsaturated in drift velocity up to high electric field and this is the same for Neon-Methane mixtures.

In figures 8 are shown the drift velocity of Xenon based mixtures. Again we see that all Xenon mixtures have saturated drift velocity at relevant electric field but Xenon-CO2 shows the highest value of the drift velocity. Xe-$CO_2$ and Xenon-Methane drift velocity measurements match very well with the simulation as shown in figures 8e and 8f.

### 6-2 Diffusion

It is evident from figure 9 that the longitudinal diffusion of argon based mixtures is typically less than 200μ/cm of drift. Argon-Methane mixtures have the least longitudinal diffusion followed by Argon-Isobutene, Argon-Methane and Argon-CO2 mixtures. Measurements and simulation of 50-50 Argon-CO2 and several percentages of methane are shown in figure 9e, the experimental points agree very well with the data.

Taking a look on helium mixtures in figure 10 we find again that ethane and isobutene with helium have the lowest longitudinal diffusion. Helium-isobutene in different percentages is shown in figure 10e with measurements and calculations of longitudinal diffusion in good agreement.

Longitudinal diffusion of Neon based mixtures in figure 11, they are very similar to the argon based mixtures. Xenon mixtures in general have a lower longitudinal diffusion coefficient as shown in figure 12; this is particularly useful for the TRT of ATLAS where they used 70% Xenon based mixtures. Simulation and measurement are in good agreement.

The transverse diffusion of all mixtures mentioned above are shown in figures (13-16) with some comparison with the measurements. The agreement is to within 5% as shown in figure 13e..

**6-3 Townsend and Attachment Coefficient.**

The Townsend coefficient is shown in figures 20-23. It increases with high electric field as expected for all argon based mixtures and the behavior is rather similar. Helium and Neon based mixtures on the other hand start multiplications at lower electric field. As an example let us take He-CO2 and He-CH4 mixtures: at 6000 V/Cm the Townsend is 5 in the former case while it is 10 for the later. Neon mixtures behave very similar. Experimental measurements for $Ar-CH_4$, $Ar-C_2H_4$ and $Ar-iC_4H_{10}$ are shown in figure 23.

Attachment coefficient is very important transport parameter to be understood as due to electron affinity gain is reduced and hence the efficiency of operation. The simulation and experimental data for a few gas mixtures are shown in figures (17-19). In figure 17b one can see the reliable estimate of the attachment coefficient can be made using the simulation package explained here. The deviation of measurement from simulation is about 15% at high electric field.

**6-4 Lorentz angle**

Operational magnetic field is very important for measurement of momentum and tracking. A large lever arm ($BL^2$) is usually employed with several measurement points in order to identify the charge and sagitta of the passing track. The CMS muon system is such an example. With the help of the simulation packages and measurements we are able to demonstrate that Lorentz effect can be precisely estimated. In figures (24-25) we present the data at magnetic field of 0.8 Tesla simply because measurement exists at this field. For argon based mixtures we see that the Lorentz angle behavior is very similar. The Lorentz angle of $Ar-CH_4$ is higher than that of $Ar-C_2H_6$. In any case the agreement with experimental data is rather impressive. He based mixtures, although have smaller value of Lorentz angle compared to Argon mixture behave very similar to each other. In figures 24c and 24d we see excellent agreement between measurement and simulation. Figures 25c and 25d show the behavior of Lorentz angle with electric field of 0.8 Tesla for mixtures based on Xenon. Experimental data are also plotted and the same plot and are in good agreement with calculations. The same is true for Neon mixtures in figures 25a and 25b.

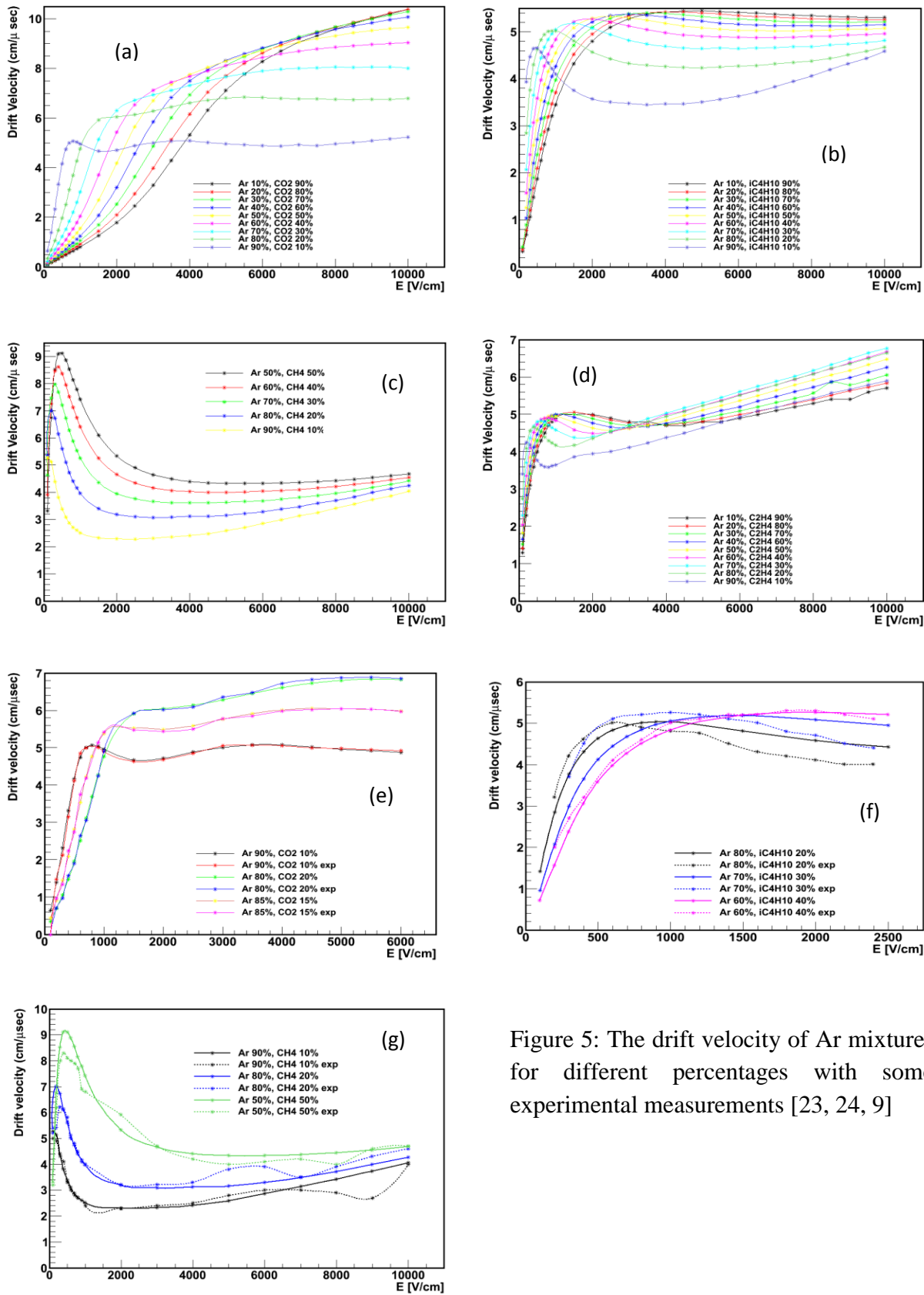

Figure 5: The drift velocity of Ar mixtures for different percentages with some experimental measurements [23, 24, 9]

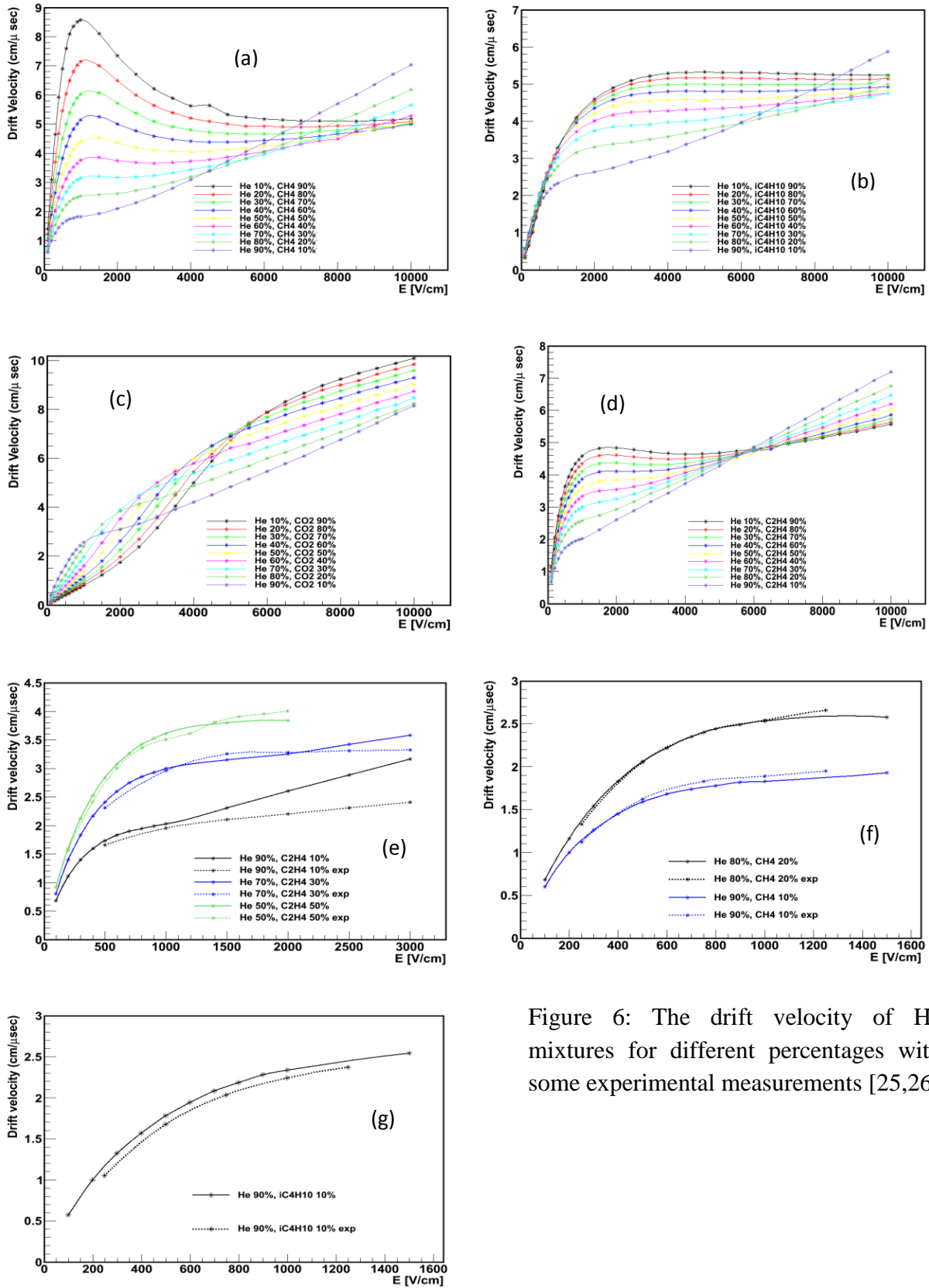

Figure 6: The drift velocity of He mixtures for different percentages with some experimental measurements [25,26]

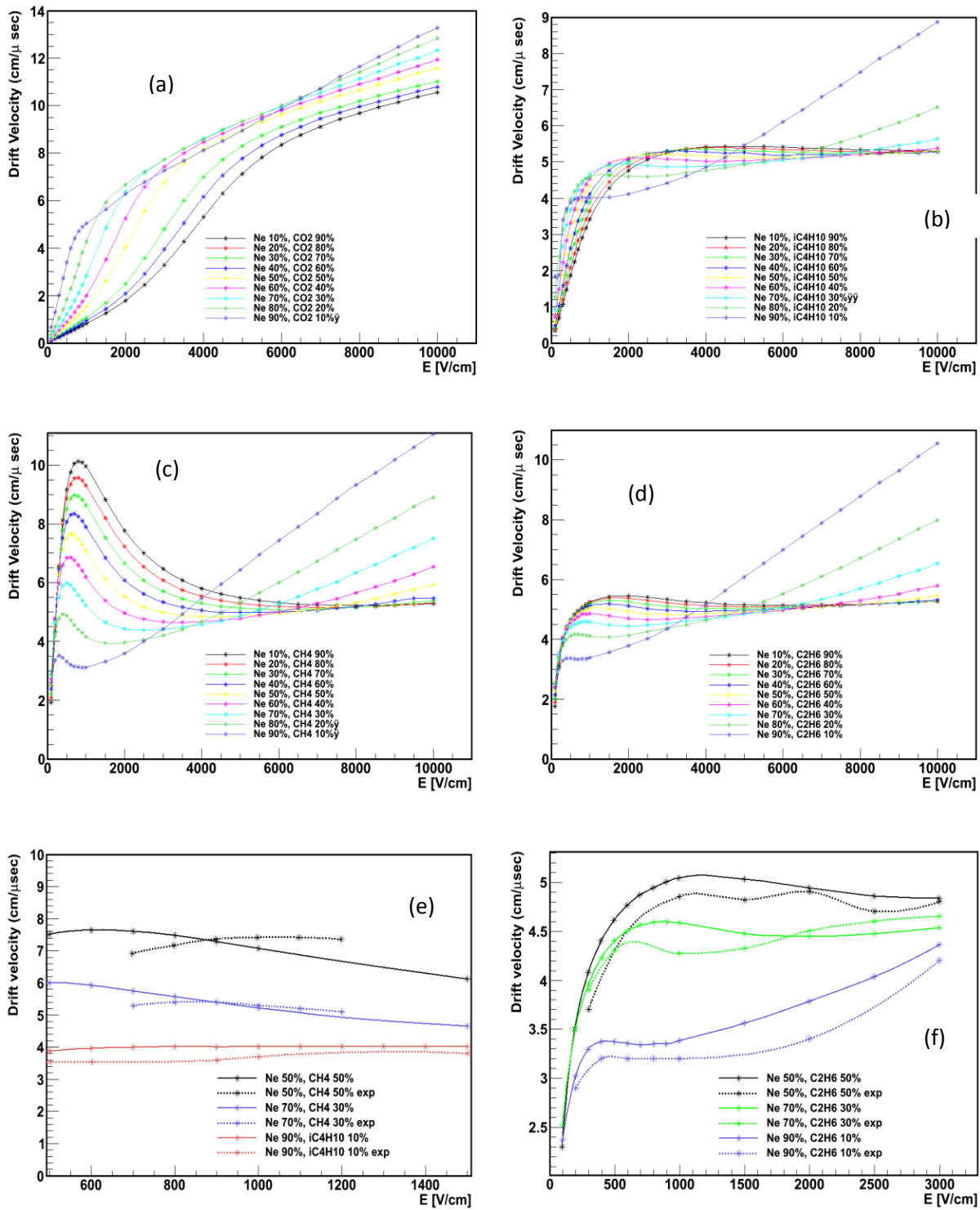

Figure 7: The drift velocity of Ne mixtures for different percentages with some measurements [24,27,28]

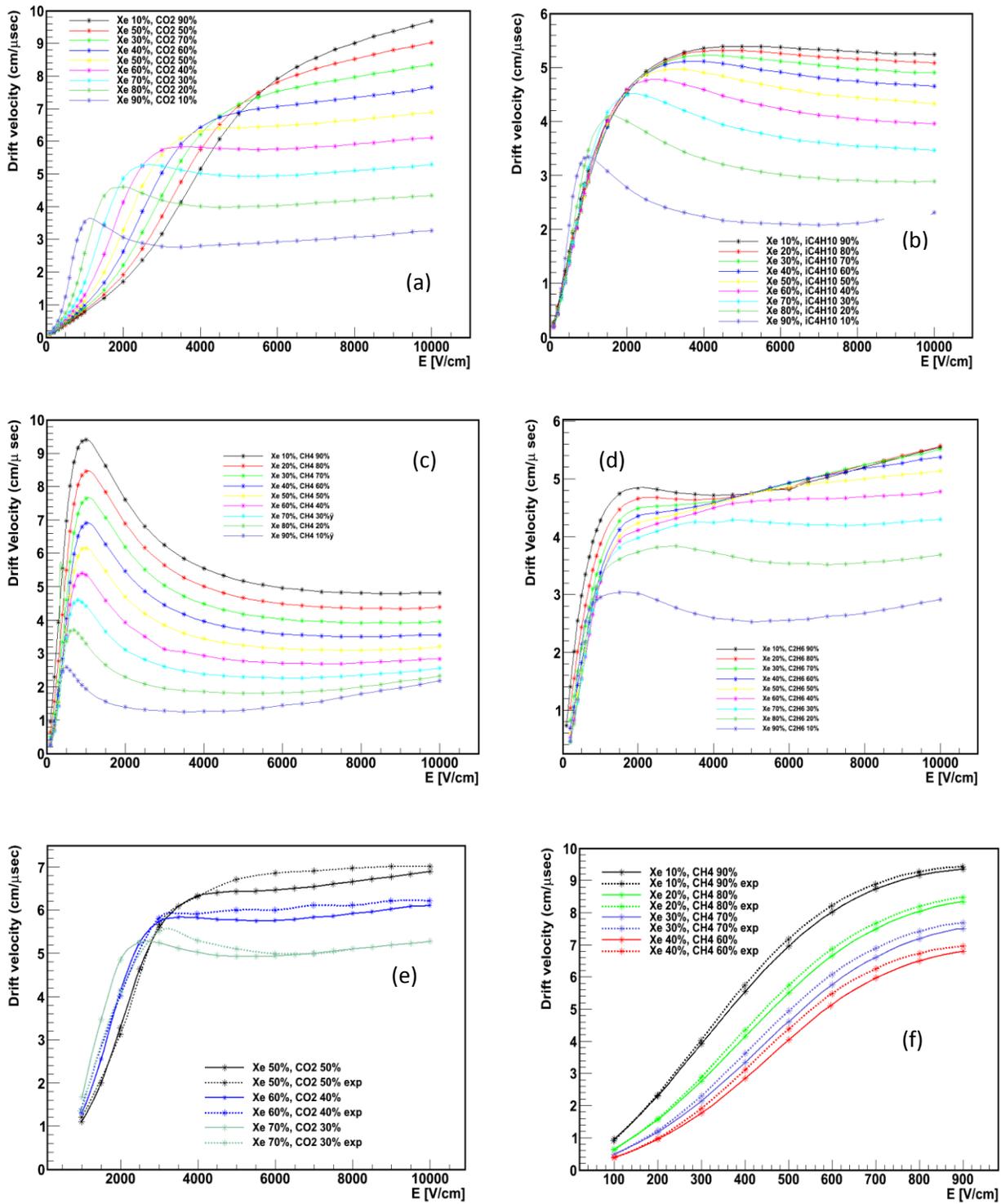

Figure 8: The drift velocity of Xe mixtures for different percentages with some measurements [29,30]

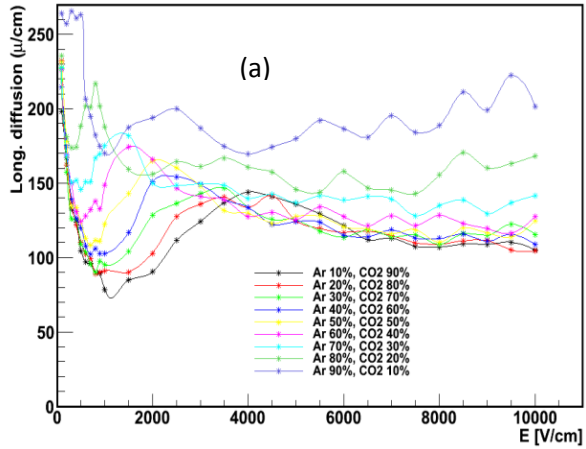
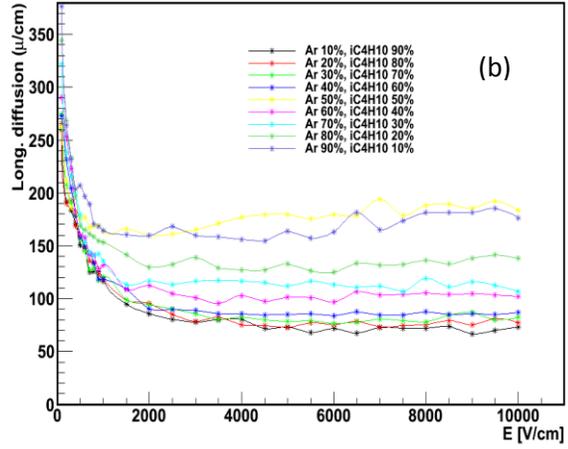
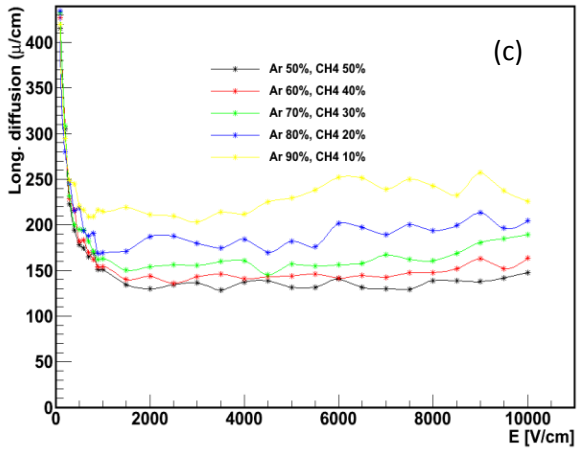
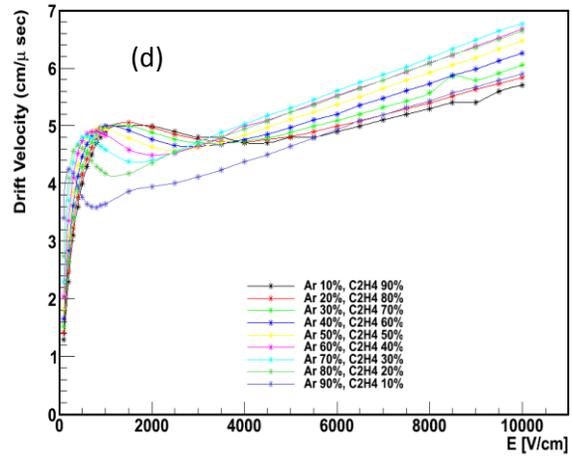
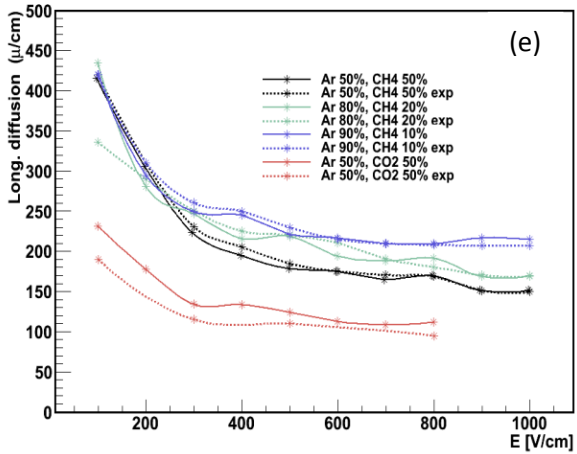

Figure 9: The longitudinal diffusion of Ar mixtures for different percentages with some measurements [31,9]

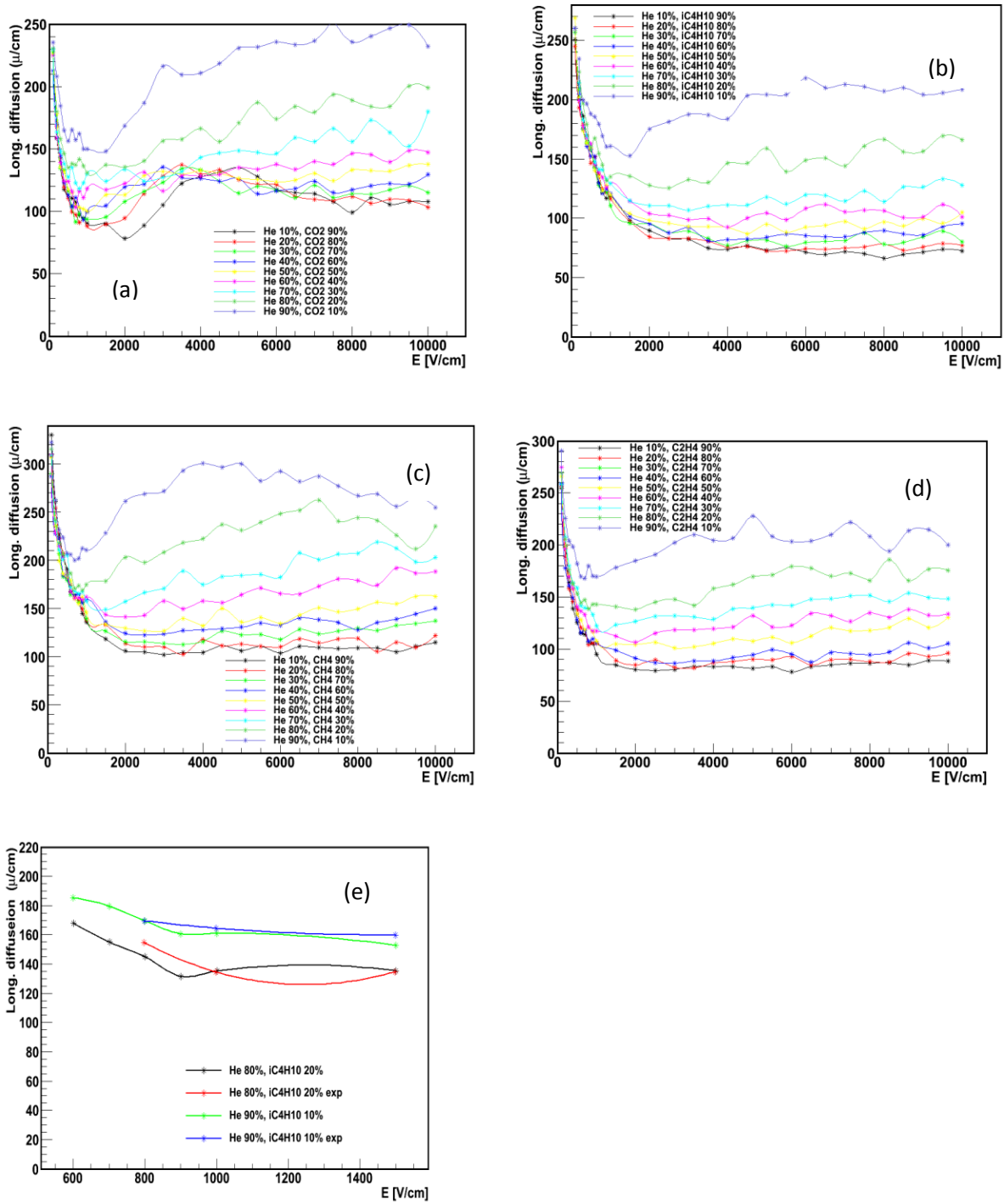

Figure 10: The longitudinal diffusion of He mixtures for different percentages with some measurements [32].

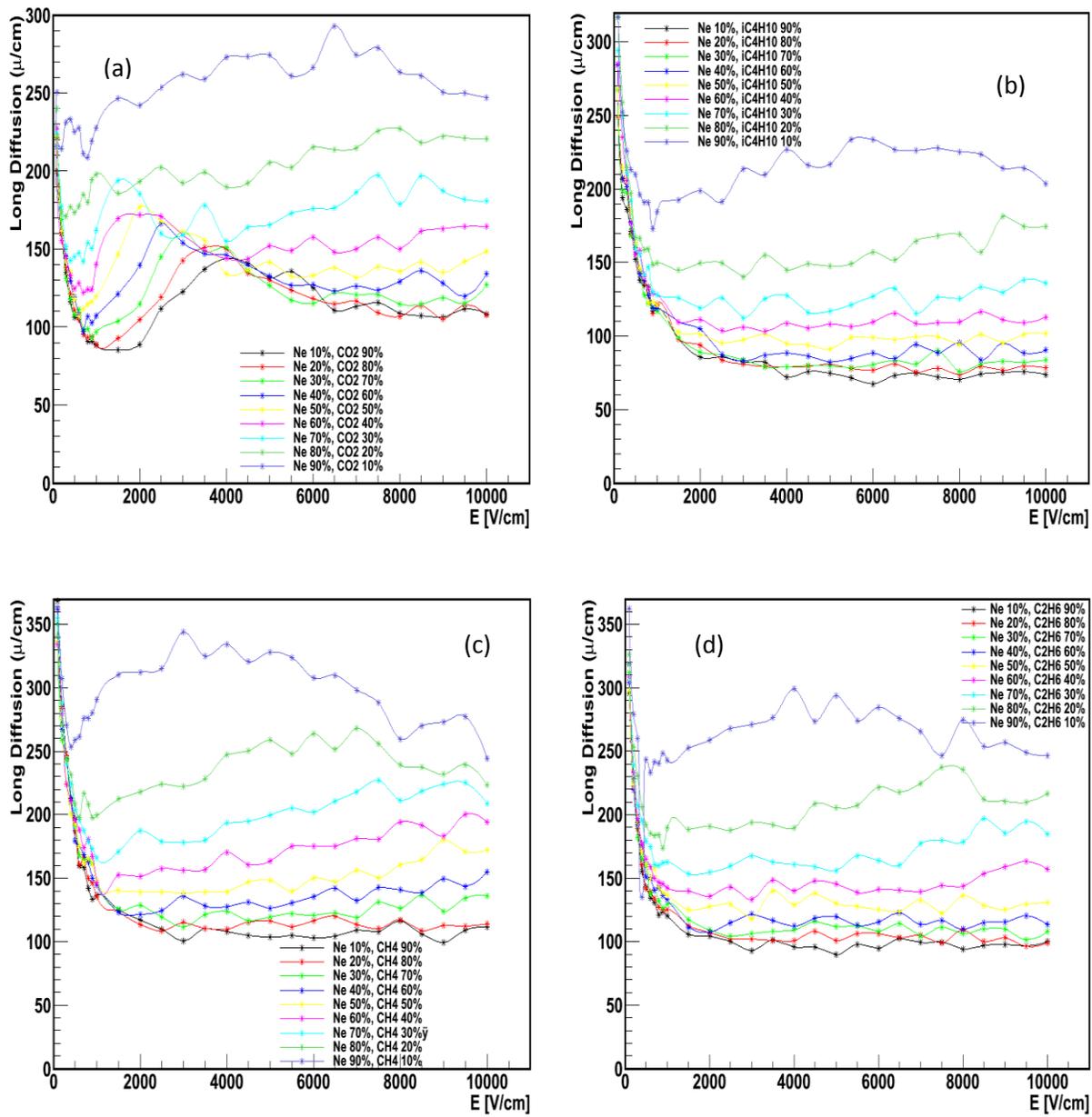

Figure 11: The longitudinal diffusion of Ne mixtures for different percentages.

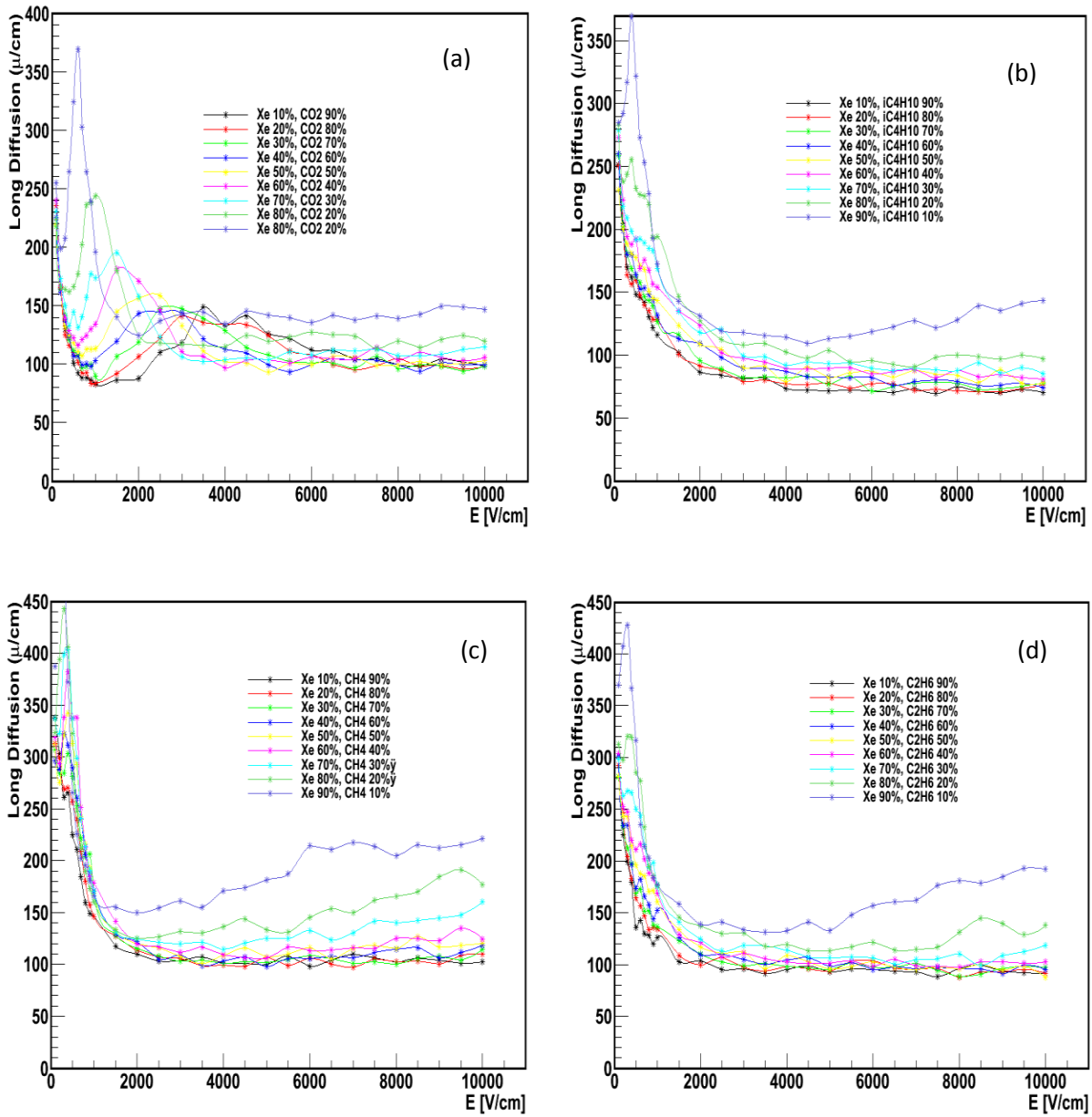

Figure 12: The longitudinal diffusion of Xe mixtures for different percentages.

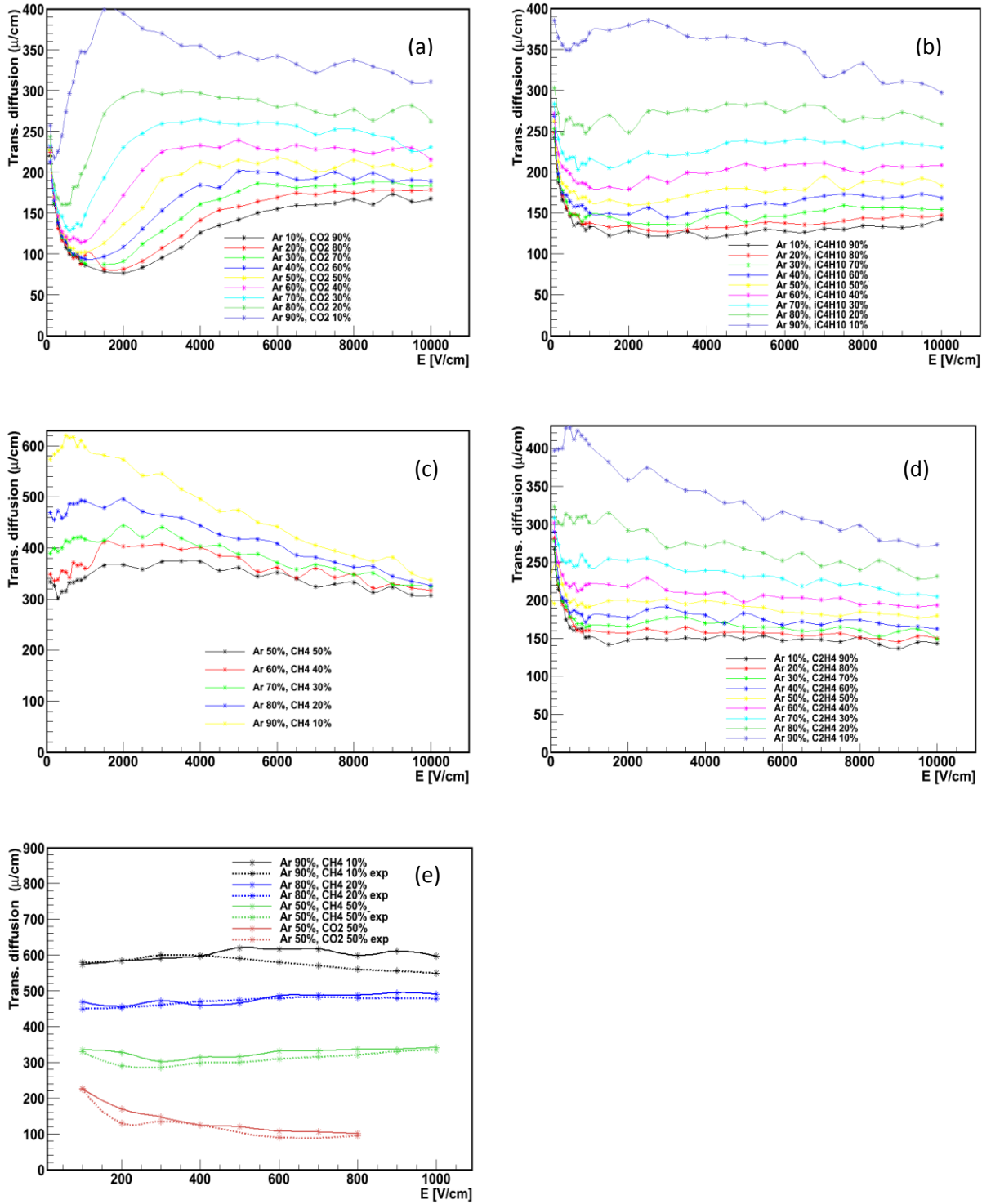

Figure 13: The transverse diffusion of Ar mixtures for different percentages with some measurements [9,31]

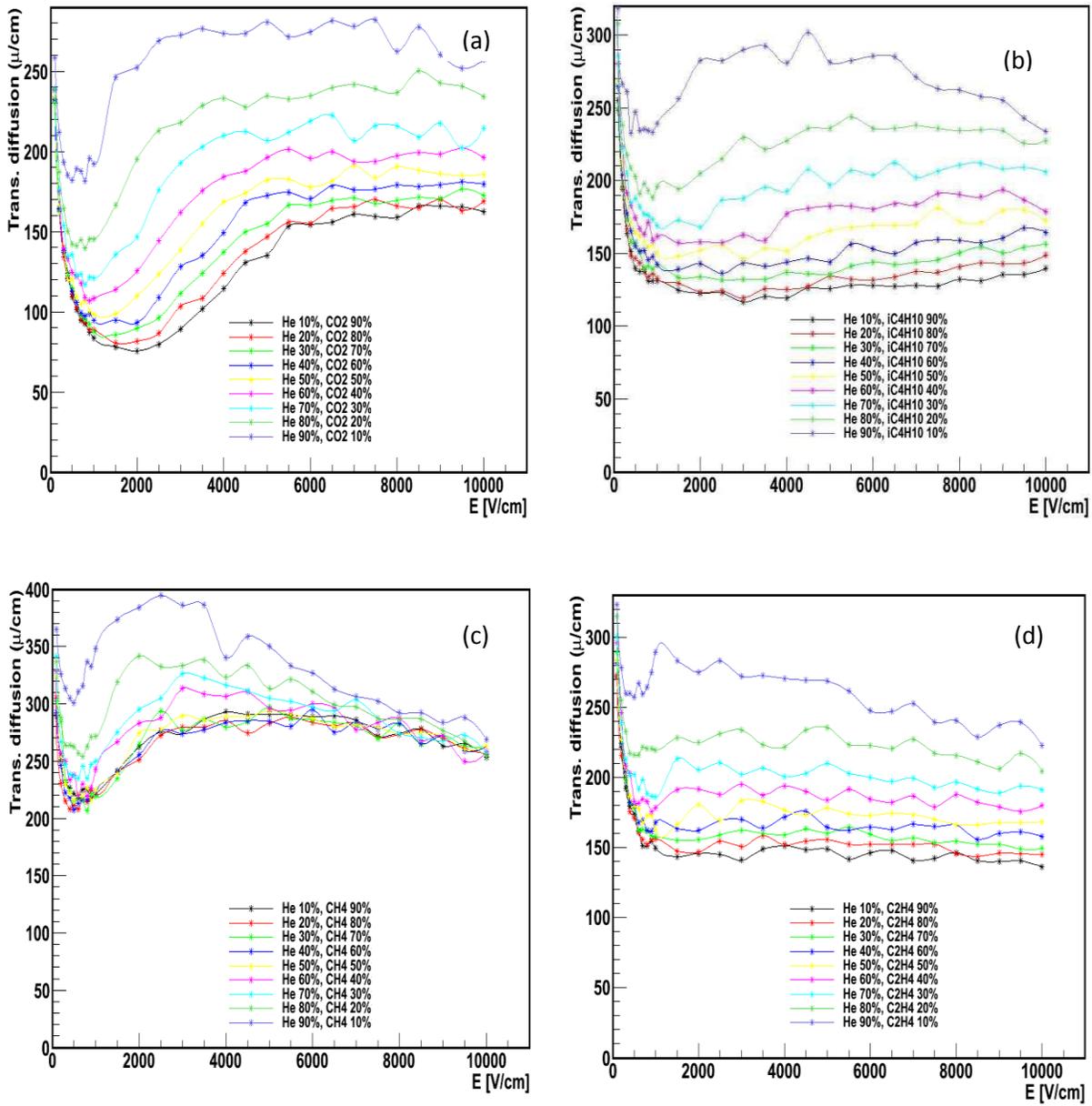

Figure 14: The transverse diffusion of He mixtures for different percentages.

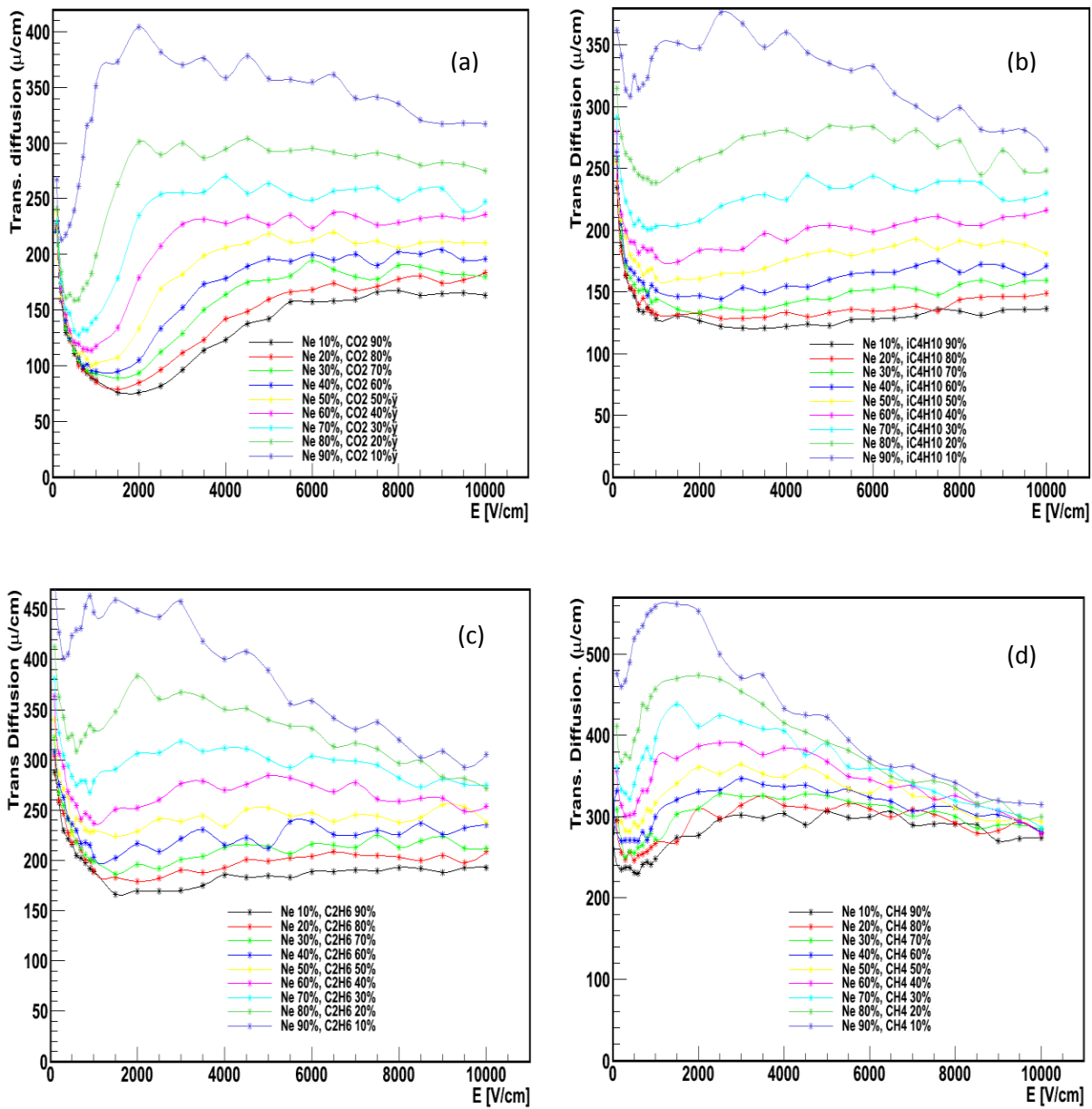

Figure 15: The transverse diffusion of Ne mixtures for different percentages.

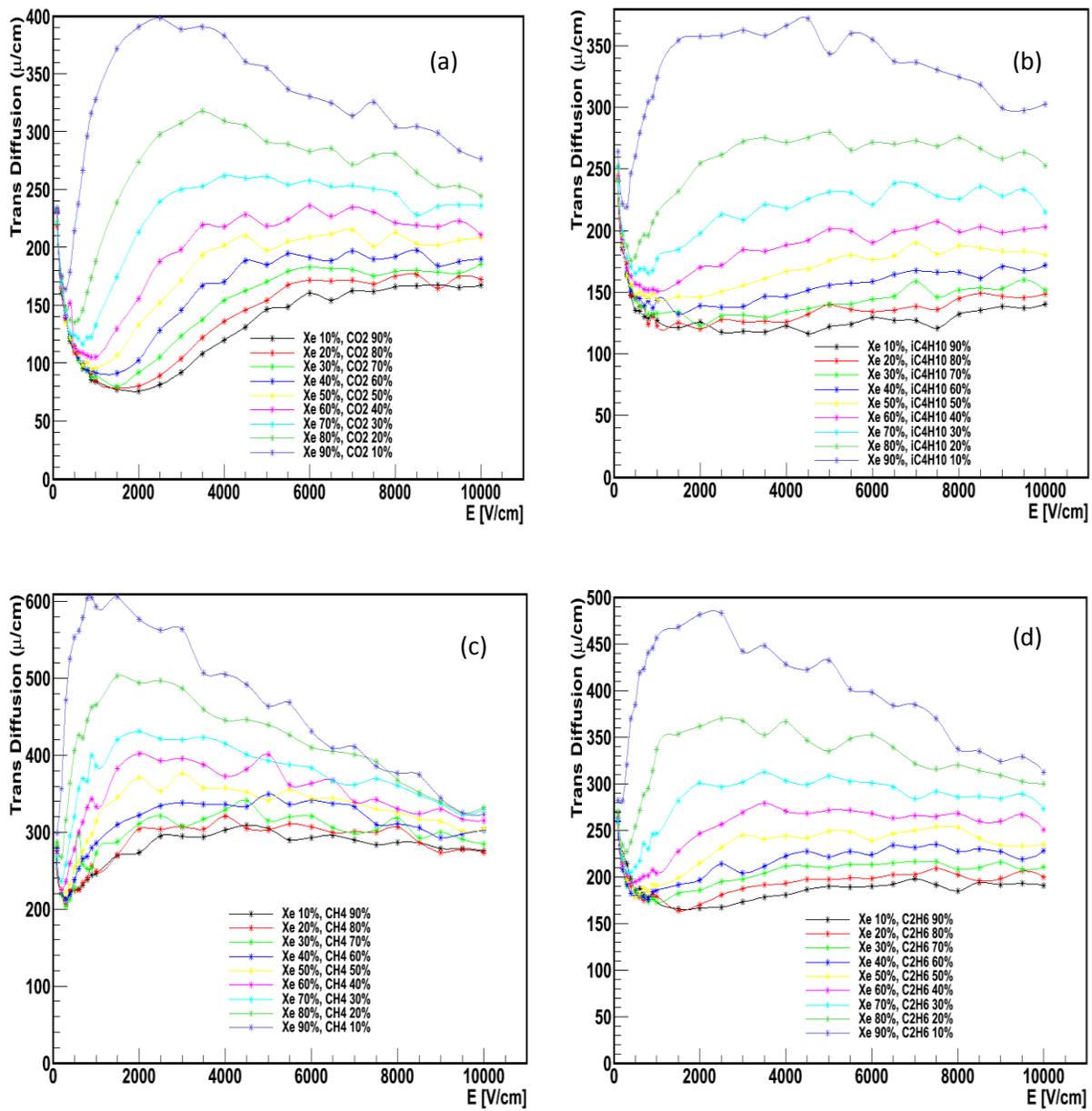

Figure 16: The transverse diffusion of Xe mixtures for different percentages.

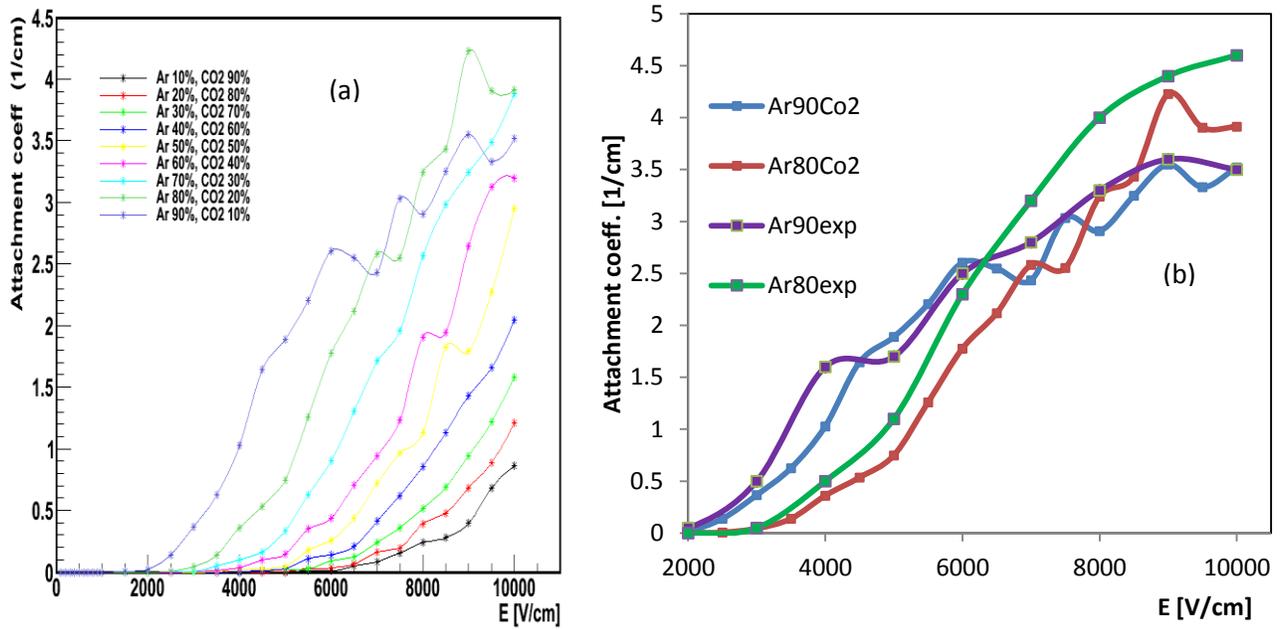

Figure 17: Attachment coefficient of Ar-CO$_2$ mixture with some measurements [33]

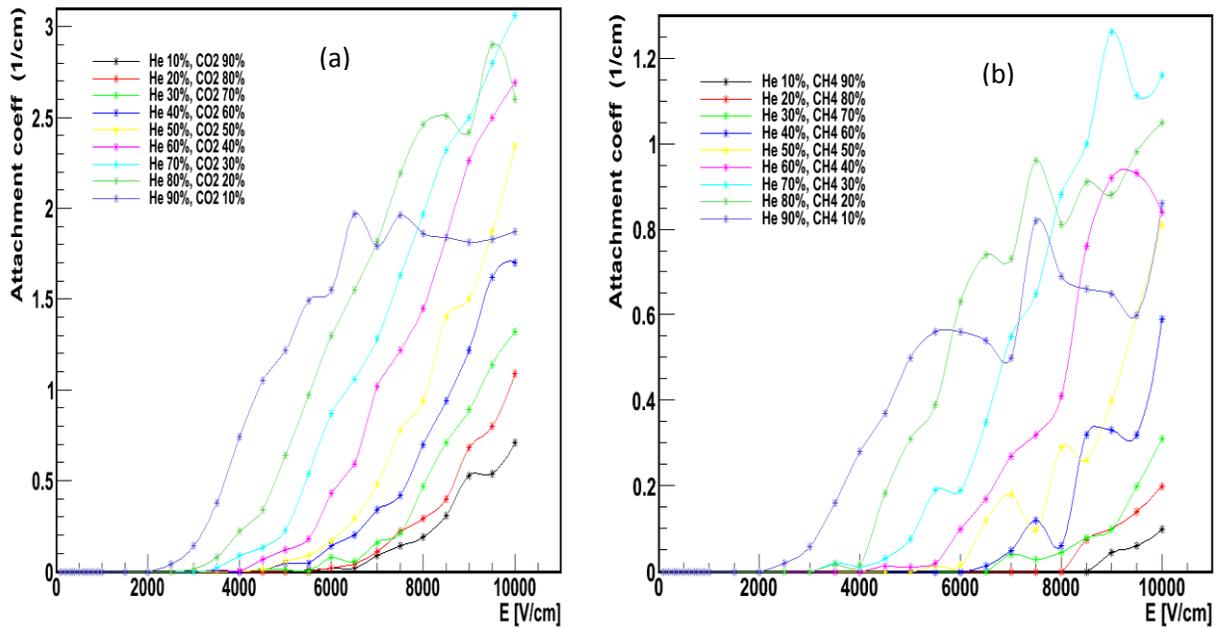

Figure 18: Attachment coefficient of He mixtures.

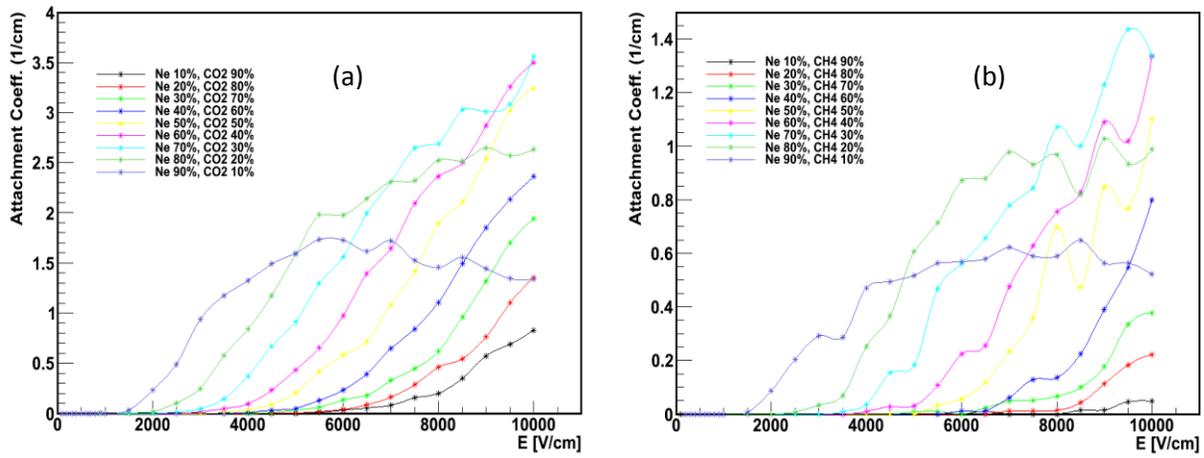

Figure 19: Attachment coefficient of Ne mixtures.

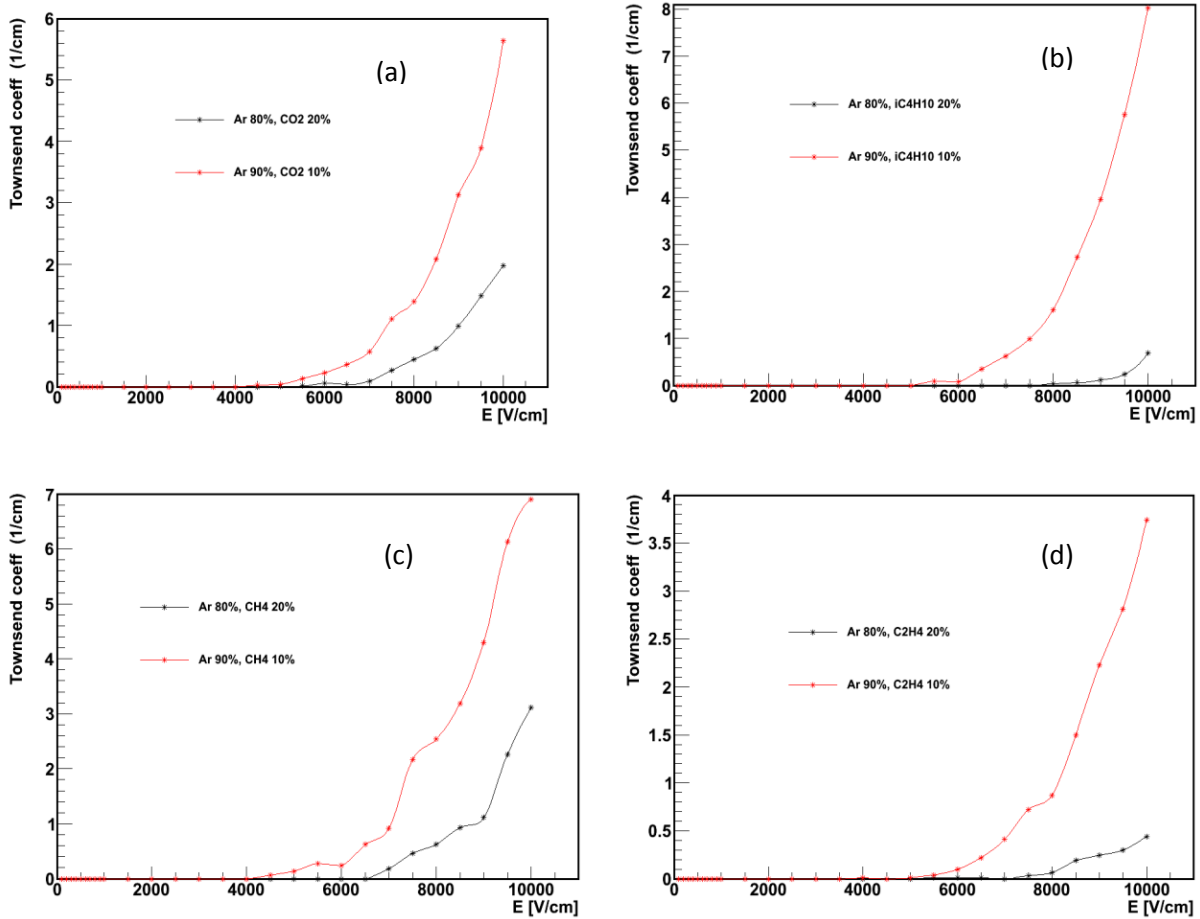

Figure 20: Townsend coefficient of Ar mixtures.

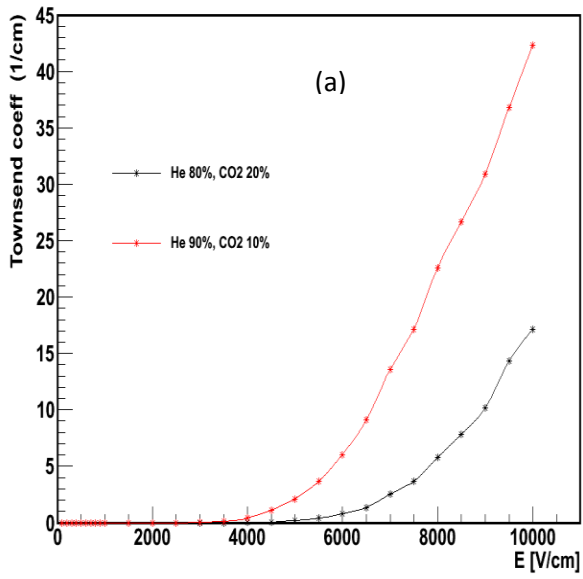
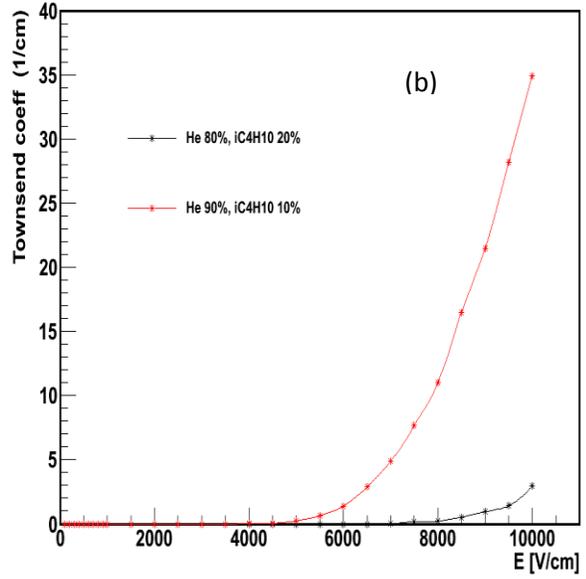
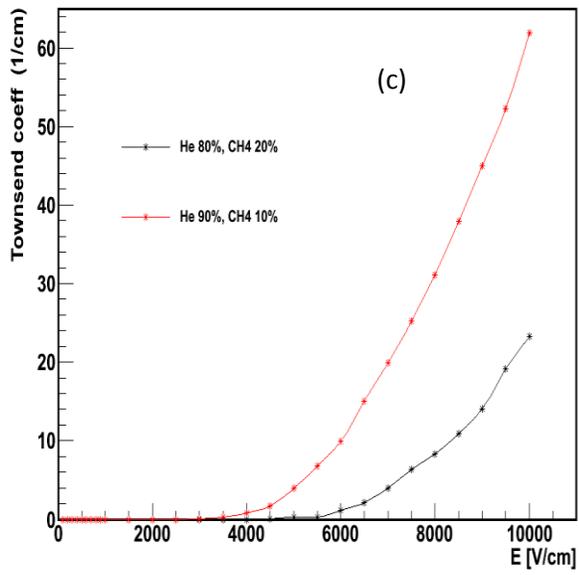
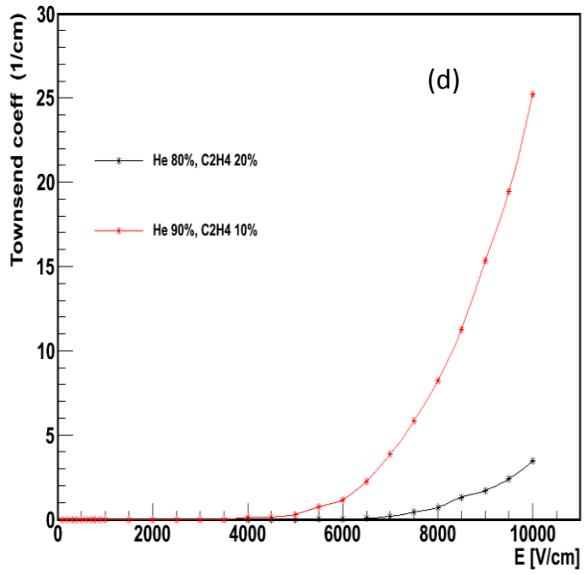

Figure 21: Townsend coefficient of He mixtures.

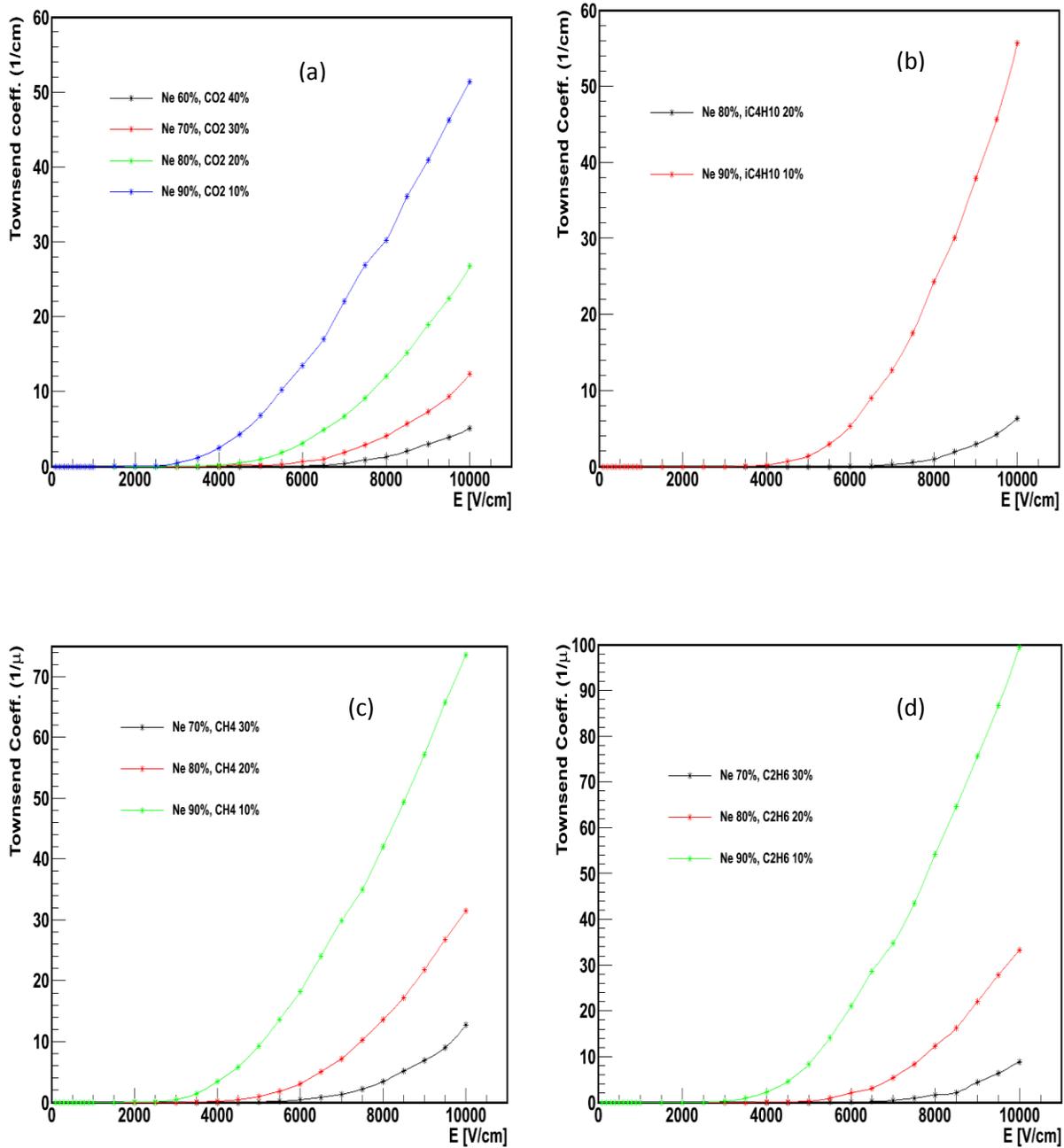

Figure 22: Townsend coefficient of Ne mixtures.

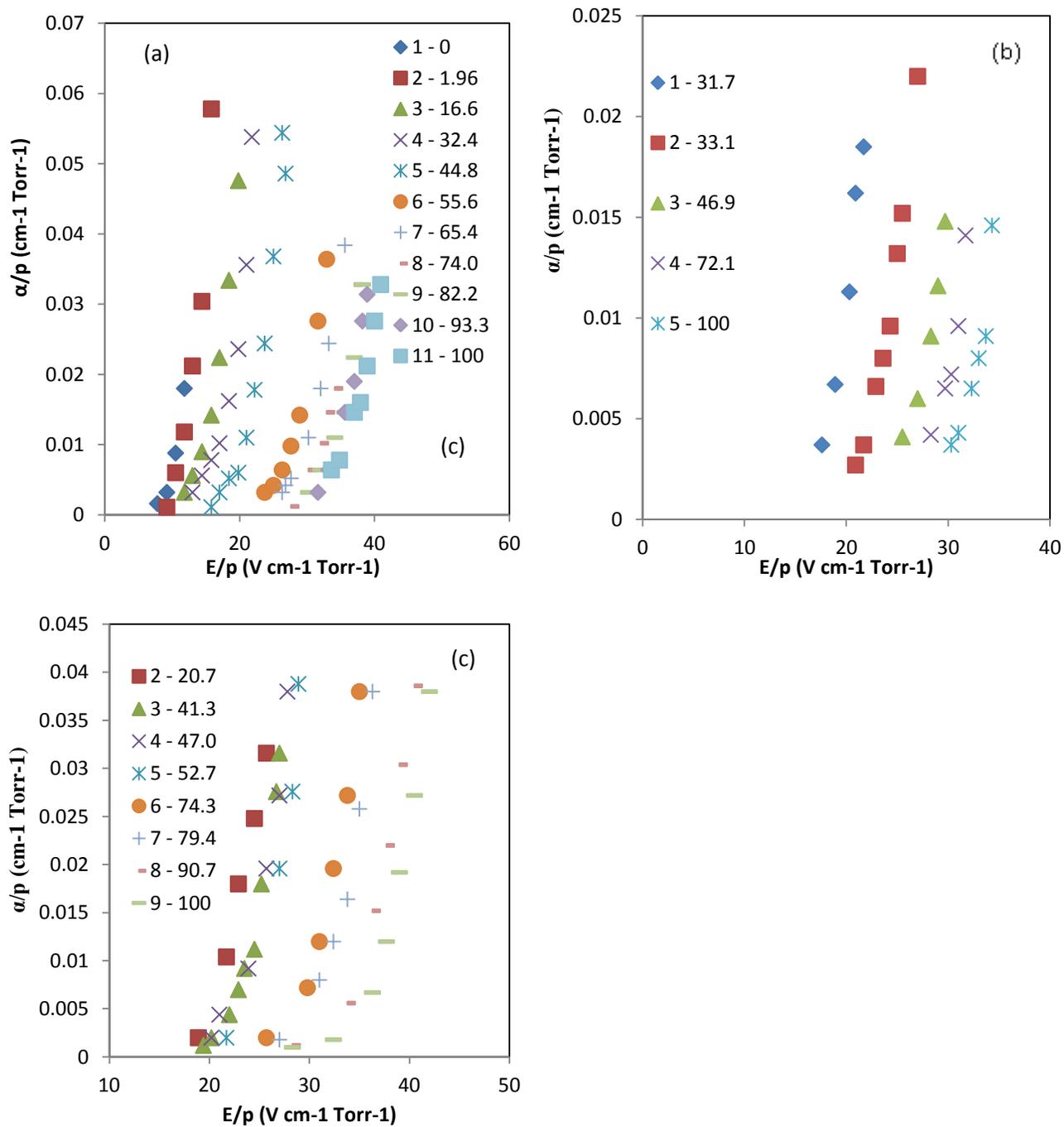

Figure 23: Measurements of Twonsend coefficient of (a) Ar-CH$_4$ , (b) Ar-C$_2$H$_4$  and (c) Ar-iC$_4$H$_{10}$ gas mixtures [34].

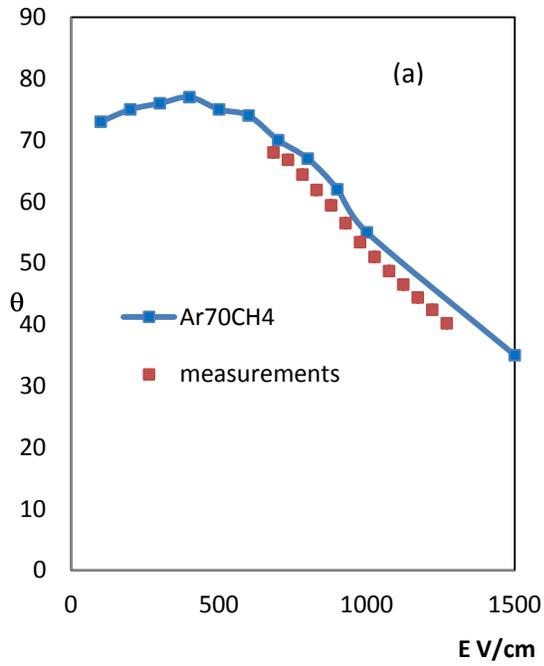
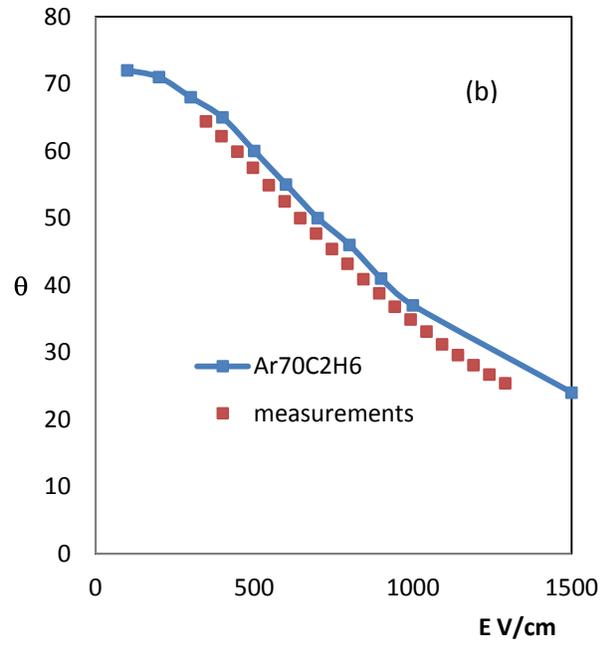

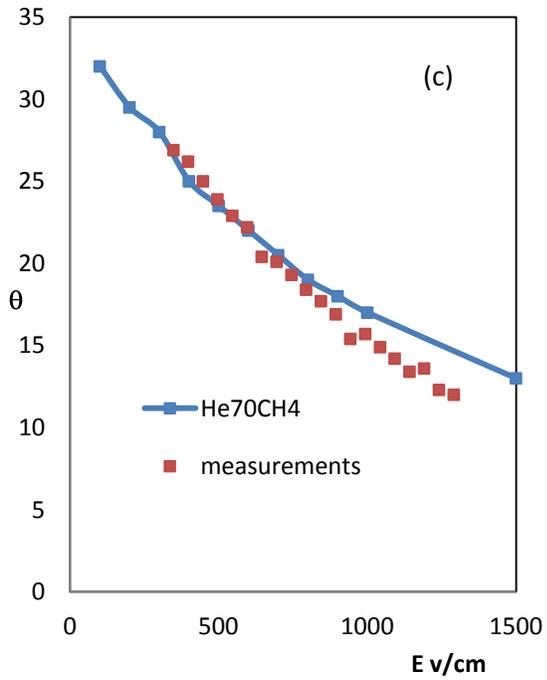
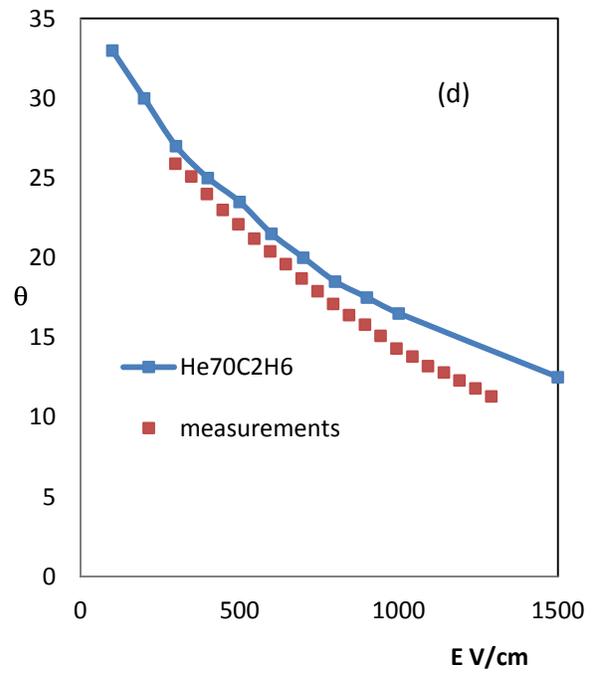

Figure 24: The Lorentz angle of (a,b) Ar mixtures and (c,d) He mixtures and some measurements at magnetic field of 0.8 Tesla.

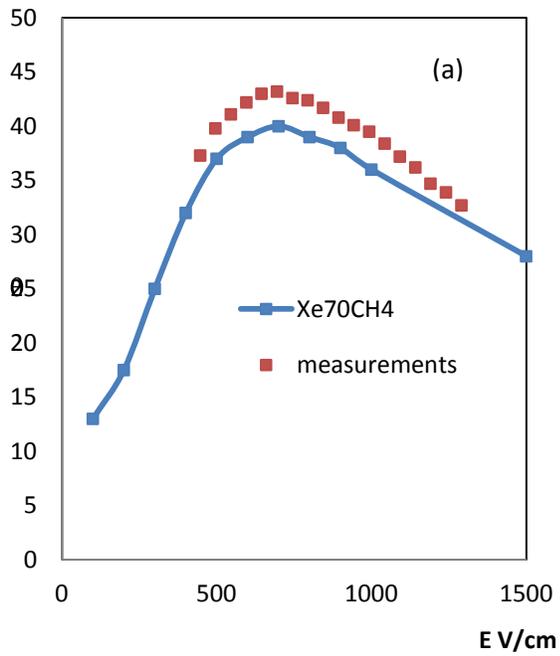
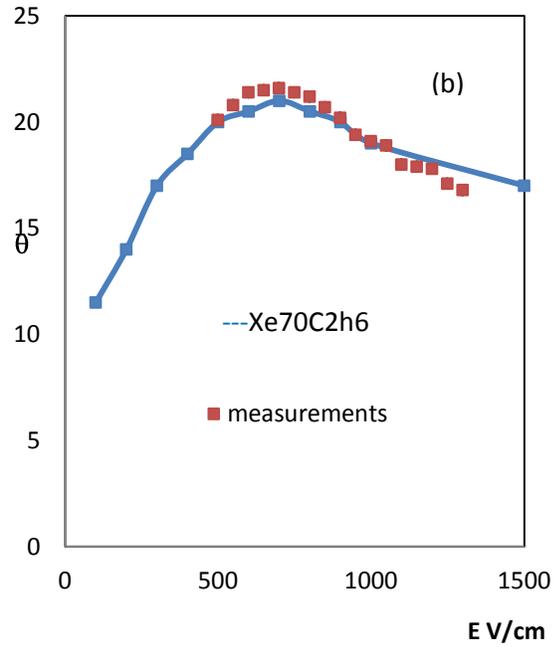

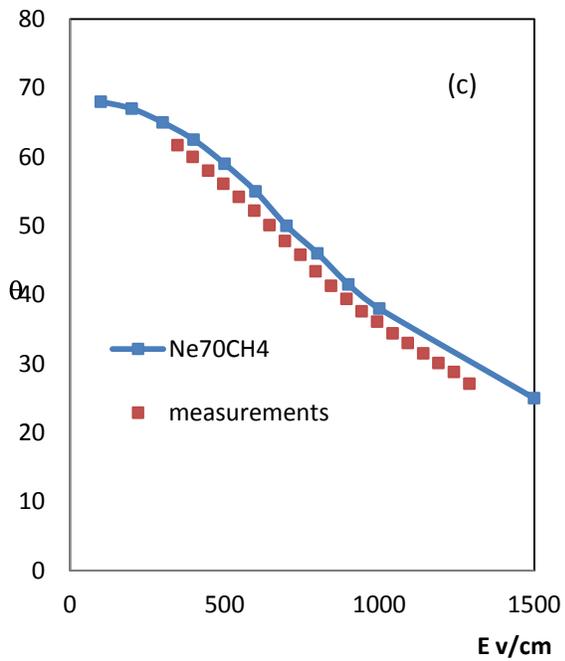
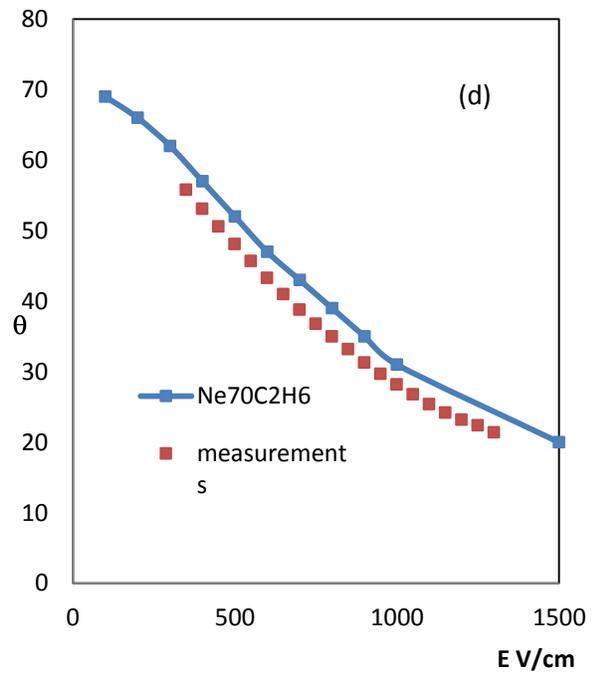

Figure 25: The Lorentz angle of (a,b) Xe mixtures and (c,d) Ne mixtures and some measurements at magnetic field of 0.8 Tesla [35].

# 7 RPC gas mixtures

The function of RPCs is trigger, so that gas mixture used should be fast gas. The gas mixture used for this purpose is Freon based (95.5% Freon, 4.2% isobutane and 0.3 SF6). Since the Freon is not an environment friendly or green gas its availability is decreasing hence alternative Freonless gas mixtures should be used. In this work, some gas mixtures were proposed and their transport parameters like drift velocity, longitudinal diffusion and transverse diffusion were estimated using Garfield software and compared to that used for RPC.

Figure 26a shows the drift velocity of some gas mixtures in addition to that used for RPC. It is shown from the figure that mixtures of combination of isobutane, carbon dioxide and argon gives the highest drift velocity. Adding small amount of Freon has the effect of lowering the drift velocity. Figure 26b shows the longitudinal diffusion coefficient of some gas mixtures with that used for RPC. In low electric field, isobutane, carbon dioxide and argon mixtures give the lowest longitudinal diffusion while in high electric field, isobutane and argon mixtures gives the lowest longitudinal diffusion. Figure 26c shows the transverse diffusion of some gas mixture and the RPC gas mixture. It is shown from the figure that isobutane, carbon dioxide and argon gas mixtures give the lowest transverse diffusion in the whole electric field range.

Figures 26d, 26e and 26f shows the Lorentz angle of some gas mixtures as a function of the electric field at magnetic field of 1, 2, 3 Tesla respectively. It is shown from these figures that for given value of the magnetic field isobutane, carbon dioxide and argon mixtures gives the lowest values of Lorentz angle. For given value of the electric field, the Lorentz angle of all gas mixtures increases as the magnetic field increases.

# 8 Conclusions

The main goal of this work is to propose an alternative gas mixture which can be used for CMS RPC and be freonless. It shown from figure 26 that isobutane, carbon dioxide and argon mixtures have the highest drift velocity, lowest longitudinal and transverse diffusion and lowest Lorentz angle so that it full the criteria of gas mixtures used for trigger purposes. So these gas mixtures have been proposed for use in the CMS RPC system and should be followed up by measurements and performance evaluation.

## Acknowledgement

I wish to express my profound gratitude and indebtedness to **Dr. Walter Van Doninck** for his invaluable help during this work.

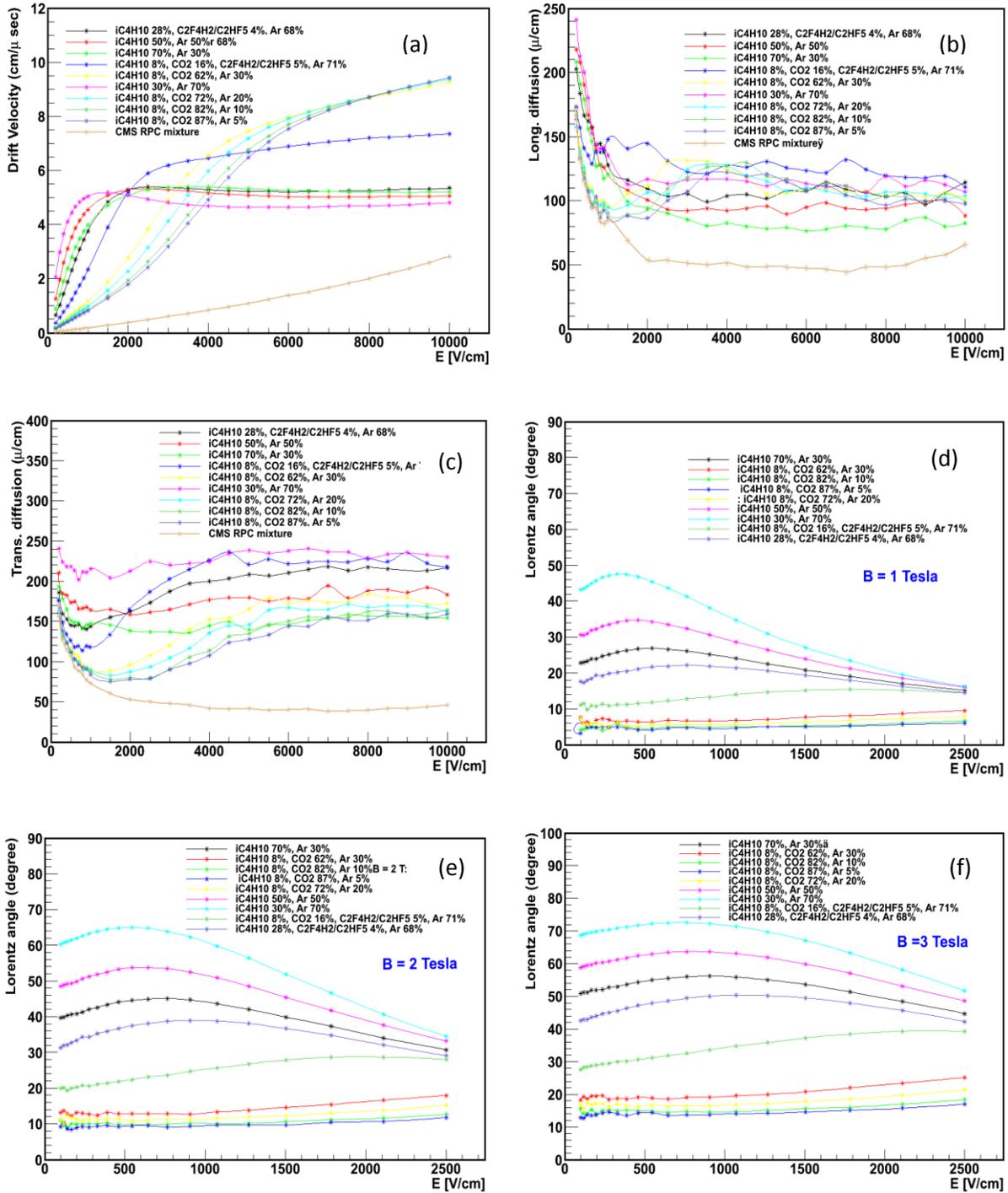

Figure 26: Transport parameters for proposed gas mixtures for CMS RPCs. (a) drift velocity ,(b,c) longitudinal and transverse diffusion and (d,e,f) the Lorentz angle at magnetic field of 1 Tesla, 2 Tesla and 3 Tesla.